\documentclass[manuscript]{aastex}

\usepackage{amsmath}
\usepackage{epstopdf}

\begin{document}

\title{MHD Shock-Clump Evolution with Self-Contained Magnetic Fields}
\author{Shule Li\\ Adam Frank and Eric G. Blackman}
\affil{Department of Physics and Astronomy, University of Rochester, Rochester, NY, 14627}
\email{shuleli@pas.rochester.edu}

\begin{abstract}
We study the interaction of strong shock waves with magnetized clumps. Previous numerical work  focused on the simplified scenario in which shocked clumps are immersed in a globally uniform magnetic field that extends through both the clump and the ambient medium. Here we consider the complementary circumstance  in which  the field is completely self-contained within the clumps. This  could arise naturally during clump formation via dynamical or thermal instabilities for example as magnetic field pinches off from the ambient medium. Using our AMR MHD code AstroBEAR, we  carry out a series of simulations with magnetized clumps that have different self-contained magnetic field configurations. We find that the 
 clump and magnetic evolution are sensitive to the fraction of magnetic field aligned with versus  perpendicular to the shock normal.   The relative strength of magnetic pressure and tension in the different field configurations allows us to analytically understand   the different cases of post-shock evolution.  We also show how turbulence and the mixing it implies depends of the initial field configuration and suggest ways in which observed shock-clump morphology may be used as a proxy for identifying internal field topologies a posteriori.
\end{abstract}


\section{Introduction}

The distribution of matter on virtually all astrophyically relevant scales is nonuniform. Within our own galaxy, matter overabundances are found in molecular clouds, and within these clouds matter further is distributed unevenly in the star-forming regions known as molecular cloud cores. Clumps of material exist on smaller scales as well. This heterogeneous distribution of matter is required, of course, for star and planet formation. On the other hand, energetic sources such as young stellar objects (YSOs), planetary nebulae (PNe), and supernovae inject kinetic energy back into their environments in the form of winds, jets, and shocks.  On larger cosmological scales galaxies are clustered implying the early evolution of the Universe involved heterogeneous or "clumpy" flows as well.  The central regions of active galaxies with their supermassive blackholes are also expected to be home to extensive regions of heterogeneous density distributions with strong incident winds and shocks. Thus understanding how the former (clumps) and latter (winds, jets, and shocks) interact remains a central problem for astrophysics.  Since dynamically significant magnetic fields are expected to thread much of the plasma in the interstellar and intergalactic medium the role of magnetic forces on shock clump interactions is also of considerable interest.

Early analytic studies of single clump/shock interaction focused on the early stages of the hydrodynamic interaction, where the solution remained amendable to linear approximations. The evolution late in time, when the behavior becomes highly nonlinear, remains intractable from a purely analytic standpoint and therefore has benefited greatly from numerical investigation -- a review of the pioneering literature may be found in \citet{kmc94} (hereafter KMC94), or \citet{pol02}. Illustrating the maturity of the field, a variety of physics has now been included in the studies. KMC94 discussed systematically the evolution of a single, adiabatic, non-magnetized, non-thermally conducting shocked clump overrun by a planar shock in axisymmetry (``2.5D''). Similar simulations were carried out in three dimensions (3D) by \citet{sto92}. The role of radiative cooling [e.g. \citet{mel02}, \citet{fra04}], smooth cloud boundaries [e.g. \citet{nak06}], and systems of clumps [e.g. \citet{pol02}] have all been studied. A similar problem involving clump-clump collisions, has also received attention [e.g. \citet{min99}, \citet{kw98}]. Most studies predominantly use an Eulerian mesh with a single- or two-fluid method to solve the inviscid Euler equations. One notable exception is \citet{pit08}, who use a ``$\kappa-\epsilon$'' model to explicitly handle the turbulent viscosity.

As the list of papers described above shows there have been many studies of hydrodynamic shock clump interactions, numerical studies focused on MHD shock-clump interactions have been fewer. Of particular note are the early studies by Mac Low [\citet{mac94}], Jones [\citet{jon96}] and Gregori [\citet{gre00}] which articulated the basic evolutionary paths of a shocked clump with an embedded magnetic field.  Further studies at higher resolution \citet{shi08} or including other physical processes such as radiative cooling \cite{fra05} or heat conduction \cite{orl05} have also been carried out.  In all these studies however the magnetic field was restricted to uniform geometries in which the field extended throughout the entire volume including both the clump, ambient and incident shocked gas.  Thus $\bf{B_o} = B_x\hat{i}+B_y\hat{j}+B_z\hat{k}$ where $(B_x,B_y,B_z)$ were constants. 

Throughout these studies the role of fields could be traced to the relative importance of components perpendicular or parallel to the shock normal. The results can be summarized as follows: (1) When the field is parallel to the shock direction, magnetic field is amplified at the head of and behind the clump.  The top of the shocked clump is streamlined but there is no significant suppression on the fragmentation of the clump even for low initial magnetic $\beta$. (2) When the magnetic field is perpendicular to the shock normal, the field wraps around the clump and becomes significantly amplified due to stretching driven by the shocked flow. In these cases the shocked clump becomes streamlined by field tension and its fragmentation via instabilities can be suppressed even for high initial $\beta$ cases. Adding radiative cooling into the simulation can further change the shocked behavior as more thin fragments and confined boundary flows, are formed [\citet{fra05}]. There are also studies in recent years focusing on the multi-physics aspect of the problem by incorporating the MHD simulations with processes like thermal diffusion, etc [\citet{orl08}].

Thus these studies with uniform fields have shown the importance of initial field geometry on the evolution of MHD shocked clumps.  The assumption of uniform fields is however an over-simplification to real environments in which clumps most likely have some internal distribution of fields which may, or may not, be isolated from the surrounding environment.  The creation of an interior field would likely be linked to ways clumps can be formed.  For example shells of magnetized gas can be swept-up via winds or blast waves. If these shells break up via dynamic modes such as the Rayleigh-Taylor (hereafter RT) or Kelvin-Helmholtz (hereafter KH) instabilities then the clumps which form are likely to develop complex internal field topologies.  While these fields may stretch into the surrounding medium reconnection can lead to topological isolation.  Numerical studies of MHD RT unstable layers relevant to supernova blast waves confirm the development of internal fields [e.g. \citet{jun95}].  Numerical and high energy density laboratory plasma experiments have also shown how collimated MHD jets can break up into clumps via kink mode instabilities [e.g. \citet{leb05}].  The clumps which form via the instability have been shown to carry complex internal fields.

Another example comes when a cold shell embedded in a hot environment attempts to evolve towards thermal equilibrium via thermal conduction. If the shell contains an initially tangled field then some of the shell material will be captured in the tangled field region and become disconnected from the background field via anisotropic thermal conduction [\citet{shu12}]. 

Thus the next level of realism in studies of MHD shock-clump interactions is the exploration of more realistic magnetic fields.  Since all studies to date have initialized their simulations with uniform fields, in this work we begin with only interior fields.  Our simulation campaign is designed to explore the question: how do more complex field topologies within the clump alter the evolution of shocked clumps.  In an effort to isolate important physical processes we choose to use relatively simple interior fields i.e. purely toroidal and purely poloidal both with different alignments to the direction of shock propagation.  While we have carried out simulations with random fields we will report the results of those studies in a subsequent paper.

In Section 2 and 3 we describe the numerical method and model. In Section 4 we report our results.  Section 5 we provide a analytic model for the evolution field energy that allows us to correctly order the different initial cases and in Section 6 we summarize and provide conclusions.

\section{MHD Equations with Radiative Cooling}
For the simulations we employed the AstroBEAR code using a 3D computational grid. The AstroBEAR code is a parallel AMR Eulerian hydrodynamics code with capabilities for MHD in two- and three-dimensions. There are several schemes of varying order available for the user.  Details on AstroBEAR may be found in \citet{cun09}, Carroll et al (2012) and at https://clover.pas.rochester.edu/trac/astrobear. While the code can treat multiple atomic, ionic and molecular species, in this work we assume the gas has solar abundance, and uniform atomic mass $\mu_A = 1.3$. 
The MHD equations that we solve numerically are as follows:
\begin{equation}
\frac{\partial \rho}{\partial t}+\nabla \cdot (\rho \bf{v}) = 0,
\end{equation}
\begin{equation}
\frac{\partial (\rho \textbf{v})}{\partial t}+\nabla \cdot [\rho \textbf{vv}+(p+\frac{B^2}{8\pi})\textbf{I}-\frac{\textbf{BB}}{4\pi}] = 0,
\end{equation}
\begin{equation}
\frac{\partial \textbf{B}}{\partial t}+\nabla \times (\textbf{v}\times \textbf{B}) = 0,
\end{equation}
\begin{equation}
\frac{\partial E}{\partial t}+\nabla \cdot [\textbf{v}(E+p+\frac{B^2}{8\pi})-\frac{\textbf{B}(\textbf{B}\cdot \textbf{v})}{8\pi}] - \Lambda(n,T)= 0,
\end{equation}
where $\rho$, $n$, $\textbf{v}$, $\textbf{B}$ and $p$ are the density, particle number density, velocity, magnetic field, and pressure, and  
 $E$ denotes the total energy density given by
\begin{equation}
E = \epsilon+p\frac{\textbf{v}\cdot \textbf{v}}{2}+\frac{\textbf{B}\cdot \textbf{B}}{8\pi},
\end{equation}
where the internal energy $\epsilon$ is given by 
\begin{equation}
\epsilon = \frac{p}{\gamma -1}
\end{equation}
and $\gamma = 5/3$. 

We denote the radiative cooling by a function of number density and temperature: $\Lambda(\rho, T)$. In our simulations, we implement the Dalgarno McCray cooling table as it is more realistic comparing to simple analytic cooling functions [\citet{dal72}]. The gas is allowed to cool to a floor temperature of $50K$ and then cooling is turned off. We define our parametere regime as ``weakly cooling" so that the region inside the clump can get cooled and hold up together but the dynamics will be mostly come from the interaction between the incoming shock and the self-contained magnetic field. This means that we require the cooling time scale behind the transmitted shock to be smaller than the clump crushing time scale by a factor of less than $10$. As we are more interested in the dynamics of the interaction mentioned above, the employment of a different cooling table or cooling floor temperature will result in similar conclusions if the ``weakly cooling" assumption is maintained.

The above MHD equations are solved with the MUSCL (Monotone Upstream-centered Schemes for Conservation Laws) primitive method with TVD (Total Variation Diminishing) preserving Runge-Kutta temporal interpolation. The resistivity is ignored in our calculation so that the dissipation of the magnetic field is numerical only.  Small artificial viscosity and diffusion are implemented in order to achieve required stability and symmetry and prevent carbuncles which can occur when strong cooling and shocks are coupled. AstroBEAR has shown excellent scaling up to $10^4$ processors (Carrol et al 2012) and the simulations presented in this paper were carried out on 1024 cores of an IBM Bluegene P machine at the University of Rochester's Center for Intergrated Computational Research.  In the simulations presented here numerical considerations with the BlueGene machine led us to turn off AMR though we carried forward AMR versions of the runs on other machines.

\section{Problem Description and Simulation Setup}
The initial conditions for the simulations presented in this paper are all based on the same clump/shock/ambient medium, conditions i.e. the clump, ambient and shock conditions are the same. The only variable we explored was the internal  magnetic field topology and strength. Our set-up for a torodial magnetic field initial condition is illustrated in Figure 01. 

\begin{figure}
\includegraphics[scale=.60]{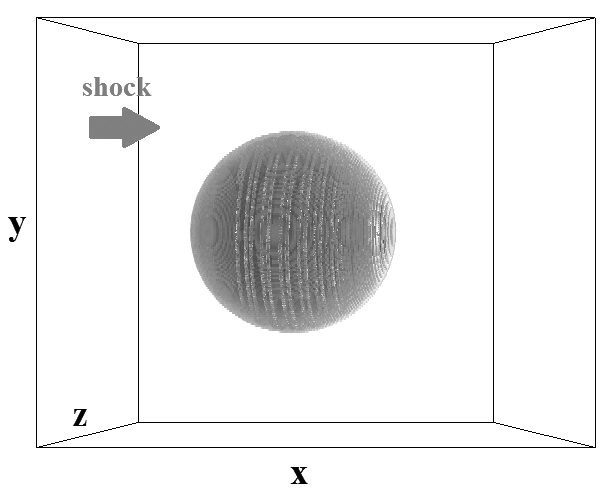}
\caption{The initial setup of the clump simulations. The actual domain is four times as long on x as on y and z. The upcoming planar shock is at the left edge of the domain, propagating rightward along the x axis. The stripes on the clump surface denote a self-contained toroidal magnetic field with its axis aligned with x axis inside the clump.}
\label{fig01}
\end{figure}

\placefigure{fig01}

We choose conditions that are astrophysical relevant with a focus on clumps occurring in interstellar environments. We note however that behavior seen in our model will scale with the appropriate dimensionless numbers.  We denote the shock speed by $v_s$, ambient sound speed by $c$, clump density by $\rho_c$, ambient density by $\rho_a$, clump thermal pressure by $P_{th}$, the self-contained magnetic field pressure by $P_B$,  clump radius by $r_c$ and radiative cooling length by $r_r$. Then as long as the Mach number $M=v_s/c$, clump density ratio $\xi = \rho_c/\rho_a$, plasma beta $\beta = P_{th}/P_{B} $ and cooling parameter $\chi = r_c/r_{r}$ are the same between two runs then the solutions should be independent of absolute scales for input parameters.

Thus we choose an ambient gas that is non-magnetized and isothermal, with a particle number density of $1 {\rm cc^{-1}}$ and a temperature of $10^4 {\rm K}$. Our clump begins with a radius of $r_c=150 {\rm a.u.}$ and is in thermal pressure equilibrium with the ambient medium. The clump has a density contrast of $\xi=100$, i.e., particle number density of $100 {\rm cc^{-1}}$ and a temperature of 100 {\rm K} . The domain is a box with dimensions $2400 {\rm a.u.} \times 60 {\rm a.u.} \times {\rm 60 a.u.}$, with an resolution of $1296 \times 324 \times 324$ , which gives $54$ cells per clump radii. We use outflow boundary conditions on the six sides of the box. We are thus able to follow the evolution for approximately 16 clump radii.  

The magnetic fields in our clumps were chosen to allow for self-contained geometeries. We use $\beta_{avg}$ to denote the ratio of thermal pressure to averaged magnetic pressure across the entire clump, that is
\begin{equation}
\beta_{avg}=\frac{P_{th}}{P_{B,avg}}
\end{equation}
where $P_{B,avg}$ denotes the average magnetic field pressure inside the clump. The detailed setup of the self-contained magnetic field is described in Appendix A.

To better characterize the initial magnetic field configuration, we use a dimensionless number $\eta$ to define the ratio of magnetic energy of the field component that is perpendicular to the shock propagation direction. If the average magnetic field energy density for the initial setup is $B_0^2/8\pi$, then the perpendicular component has an average magnetic field energy denisty of $\eta B_0^2/8\pi$, while the parallel component has an average magnetic field energy density of $(1-\eta)B_0^2/8\pi$. $\eta$ for different initial magnetic field setup is summarized in Table 01.

Throughout the paper, we use $\beta_{avg}$ as a measure of dynamical importance of the self-contained magnetic field, and investigate the shocked behavior of situations where the self-contained field is either strong or weak. We will refer to the simulations with $\beta_{avg} = 0.25$ as "strong" field cases and those with $\beta_{avg} = 1.0$ as "weak" field cases throughout the paper. The orientation of the magnetic field relative to the incident shock is another critical parameter.  This was already seen in the uniform field simulations described in the introduction. In our simulations, we focus on the cases when the self-contained magnetic field is either purely poloidal or purely toroidal. For these fiield configurations which possess an axial symmetry, it will be the orientation of the field axis $\bf{b}$ to the shock normal $\bf{n}$ which matters.  For each configuration we run both parallel $\bf {b\cdot n} = 1$ and perpendicular cases $\bf{b\cdot n} = 0$.  The complete set of runs presented in this study are described and coded in Table 01 and these orientations are presented visually in Figure 02.

\placetable{tab01}

\begin{figure}
\includegraphics[scale=.30]{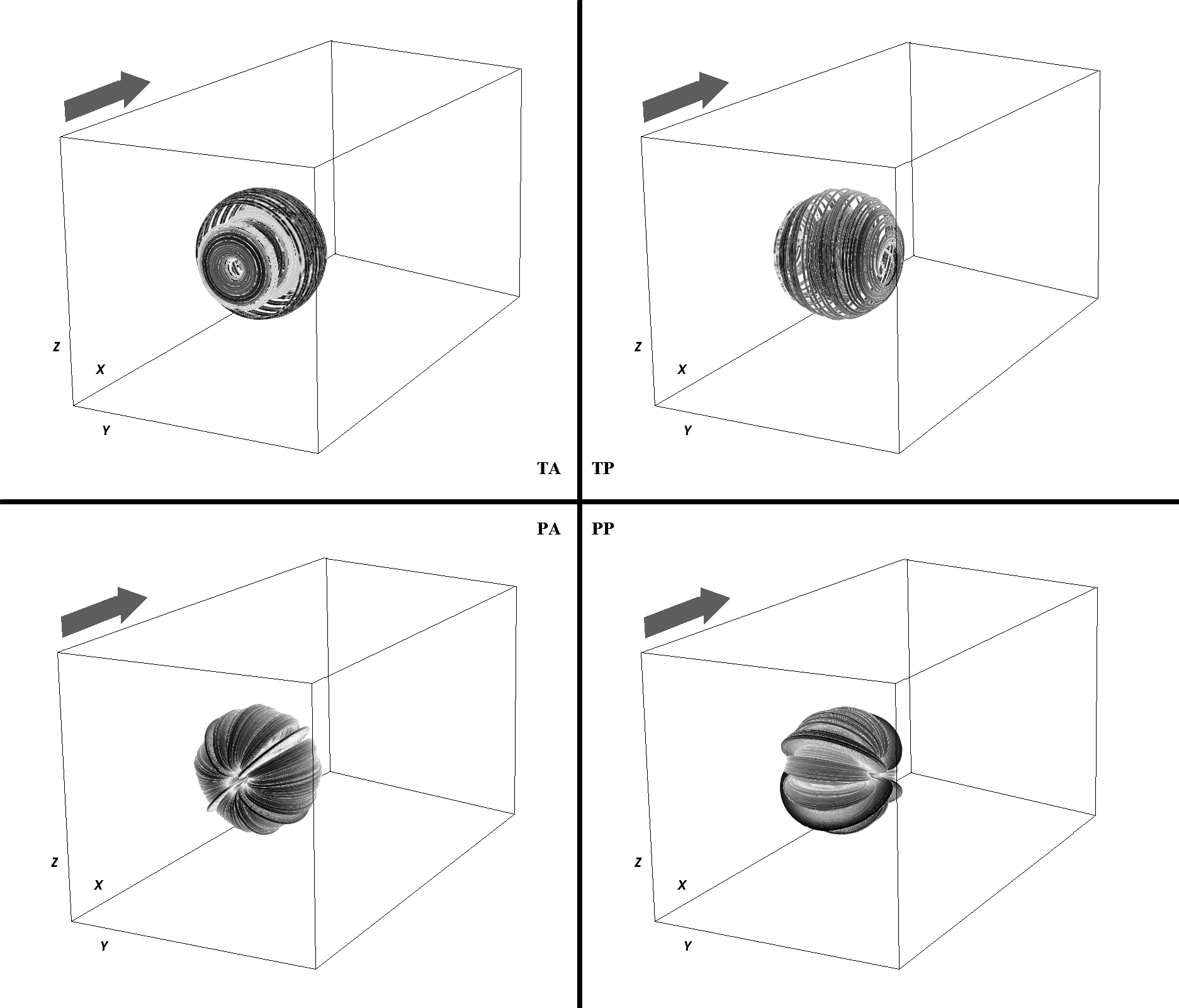}
\caption{The initial setup of the clump magnetic field. The actual domain is four times as long on x as on y and z. The first letter denotes the field configuration: T for toroidal only; P for poloidal only. The second letter denotes the field orientation with respect to the shock propagation direction: A for aligned; P for perpendicular. The blue arrow denotes the shock direction.}
\label{fig02}
\end{figure}

\placefigure{fig02}

We do not begin our simulations in a force free state as it is not clear that this is the most generic astrophysical situation.  Clumps created in dynamic environments subject to repeated incident flows may not have time to relax to force free conditions.  Thus we expect the clump will be deformed by the self-contained field on the time scale of
\begin{equation}
\tau_{B}=\frac{r_{c}}{u_A} \approx 436 {\rm yrs},
\end{equation}
where $u_A$ is the Alfven speed of the self-contained field calculated from the average magnetic energy density inside the clump. In our simulations the clump evolution driven by the shock is always faster than or comparable to this timescale as we discuss below. 

The incoming shock has a Mach number $M =10$ which puts our simulations in the strong shock regime (KMC94). To understand the role of the magnetic fields we identify the clump crushing time scale as
\begin{equation}
\tau_{cc}=\frac{\sqrt{\chi}r_{c}}{v_s} \approx 95 {\rm yrs}.
\end{equation}
Thus $\tau_{cc} < \tau_B$ and we expect that the strong shock dynamics driven by the transmitted wave propagating into the clump will dominate over any relaxation driven effects from the internal magnetic field.  To confirm this we also define energy parameters of the shock clump interaction where $\sigma_{th} = K_s / E_{th}$ and $\sigma_{B} = K_{s} / E_{B}$. These are ratios between shock kinetic energy density $\propto \rho_s v_s^2$ and the thermal or average magnetic energy density contained in the clump, respectively.  From parameters for our simulation we then have $\sigma_{th} \approx 222$ and $\sigma_{B} \approx 50$. Thus, although the clump is initially magnetically dominated, the shock has higher energy densities than either the thermal or magnetic energy contained inside the clump. Given these conditions and our choice of $\tau_{cc} < \tau_{B}$ we expect that most of the simulation evolution will driven by the shock and not internal relaxation.

We note that the cooling time scale for the transmitted shock $\tau_{r} = E_t/\dot{E_t} = kT_p/n \Lambda$ is below the clump crushing time to ensure noticeable cooling and is given by 
\begin{equation}
\tau_{r} \approx 48 {\rm yrs} \ll \tau_{cc}.
\end{equation}
Therefore we are in the regime of ``weakly cooling" inside the clump where the magnetic energy is concentrated, i.e., for the transmitted shock, the ratio of cooling time against crushing time $\chi = \tau_{r}/\tau_{cc} < 1$. The cooling length scale can be calculated as:
\begin{equation}
l_r = v_{ps} \tau_{r}
\end{equation}
where $v_{ps}$ is the post-shock sound speed:
\begin{equation}
v_{ps} = \sqrt{\frac{\gamma k_B T_{ps}}{m_A}}
\end{equation}
From the above equations, we can calculate the ratio of the clump radius to the cooling length behind the transmitted shock:
\begin{equation}
chi_* = r_c/l_r \approx 5.64
\end{equation}
Therefore one clump radius contains $5$ cooling length scales. The bow shock in our simulations has a cooling time that is longer than the evolutionary timescale of the flow and remains adiabatic in its dynamics. Notice that although the situation we consider here is freely scalable, the condition ``weakly cooling" should always be satisfied.  Since the cooling length scale does not depend on the size of the clump, it can become extremely small comparing to the clump radius when the scale length is increased and thus become a dominating process after applying such a scaling. 

We run the simulation from time $t=0$ to time $t \approx 333 {\rm yrs}$ or  $t \approx 3.5\tau_{cc}$. We will use the clump crushing time $\tau_{cc}$ as our unit of time throughout the rest of the paper.  

\section{Simulation Results}
\subsection{Shocked clumps with a self-contained strong ordered field}
We begin with the simulations in which the internal self-contained magnetic field is relatively strong ($\beta_{avg}=0.25$). Recall in what follows that the incident shock kinetic energy is dominant in the initial interaction even though the clump is magnetically dominated in terms of its own initial configuration. Figure 03 shows case TAS: i.e. the internal magnetic field is toroidal and aligned with the shock normal. Panels run from top to bottom and correspond to different evolutionary times: $t = (\tau_{cc}, 2\tau_{cc}, 3.5\tau_{cc})$. 

At early times, $t\le\tau_{cc}$, the shocked clump evolution appears similar to that of the unmagnetized case (not shown). The usual pair of shocks form: a bow shock facing into the incoming flow and a transmitted shock which propagates into the clump.  Note that the transmitted shock in our simulations is radiative meaning that thermal energy gained at the shock transition is quickly radiated away. With the loss of thermal pressure support the shock collapses back towards the contact discontinuity. In this regime shock regions becomes thin and post-shock densities are high [\citet{yir10}]. In our simulations, only the bow shock cools effectively which is evident at the thin boundary flows. 

The effect of the toriodal field becomes particularly apparent in the morphology after a crushing time.  At the middle frame in Figure 03 ($2\tau_{cc}$) we see the clump collapsing towards the symmetry axis due to the pinch by the toroidal magnetic field. This behavior is in contrast to the hydrodynamic or MHD adiabatic case with parallel fields in which the shocked clump material expands laterally and is then torn apart by RT instabilities.  Even in radiative hydrodynamic cases the shocks tend to flatten the clump which then break up into clumps [\citet{yir10}].  Only in uniform perpendicular field cases do we see situations where the flow becomes shielded from RT instabilities.  The internal toroidal field simulations show something different entirely however.  Here the tension force from the compressed internal toroidal field is strong enough to suppresses the lateral expansion. This inward directed tension controls the subsequent evolution.

The ongoing compression within the clump driven by the tension of the torodial field restricts the downstream flow.  Thus only a limited turbulence wake forms. The compression of the clump and downstream flow into a narrow cone continues at later times as can be seen in the frame corresponding to  $t=3.5\tau_{cc}$.  By this time shocked clump has become compressed into a very narrow conical feature resembling the "nose cone" observed in the MHD jet simulations [e.g. \citet{ada98}, \citet{lin89}]. The development of an dense streamlined clump by the end of the simulations indicates that for these configurations the long term evolution will be simply slow erosion of the clump without significant fragmentation.

When the toroidal axis is perpendicular to the shock normal however the evolution is quite different.  In Figure 04 we show 3 snapshots of density for run TPS.   In this case the field is attempting to pinch the clump onto z axis (a compression "inward" towards the clump axis along the $x$ and $y$ directions). The shock however only produces a compression along the $x$ axis. The differential forces on the clump do yield on transient period of flattening as is seen in both hydrodynamic and uniform field MHD simulations.  However the presence of the internal toroidal fields alters the internal distribution of stresses.  The result is a differential aerodynamical resistance to the flow over the clump as it becomes immersed in the post-shock region. Note that the magnetized clump is easier to distort along $z$ axis compared to $y$ axis where tension forces are at work.  Thus at $t=\tau_{cc}$ we see the clump becoming ellipsoidal or football shaped. The structural coherence that the tension force provides in $y$ direction during the compression phase continues to shape the subsequent flow evolution.  By $t=2\tau_{cc}$ oblate clump which continues to be eroded by the incoming wind begins developing a concave morphology along the $z$ axis.  The subsequent formation of a "banana" shaped configuration tilts the field along the body of the clump shifting the position of the local toroidal axis relative to the incident flow.  Thus the clump begins to fragment mostly along the $z$ axis because of a lack of field tension in this direction. In addition a ring-like feature  develops along the outer extent of the clump where the field is initially concentrated.  By the end of the simulation, the clump has fragmented along the z axis from  erosion and cooling, and evolves to an array of cold, magnetized "clumplets". 

Note that the perpendicular toroidal case produces a turbulent wake that occupies a much larger volume than the parallelly oriented case.  As we will see the development of such an extended wake is well-correlated with the degree of mixing between clump and ambient medium.

\begin{figure}
\includegraphics[scale=.25]{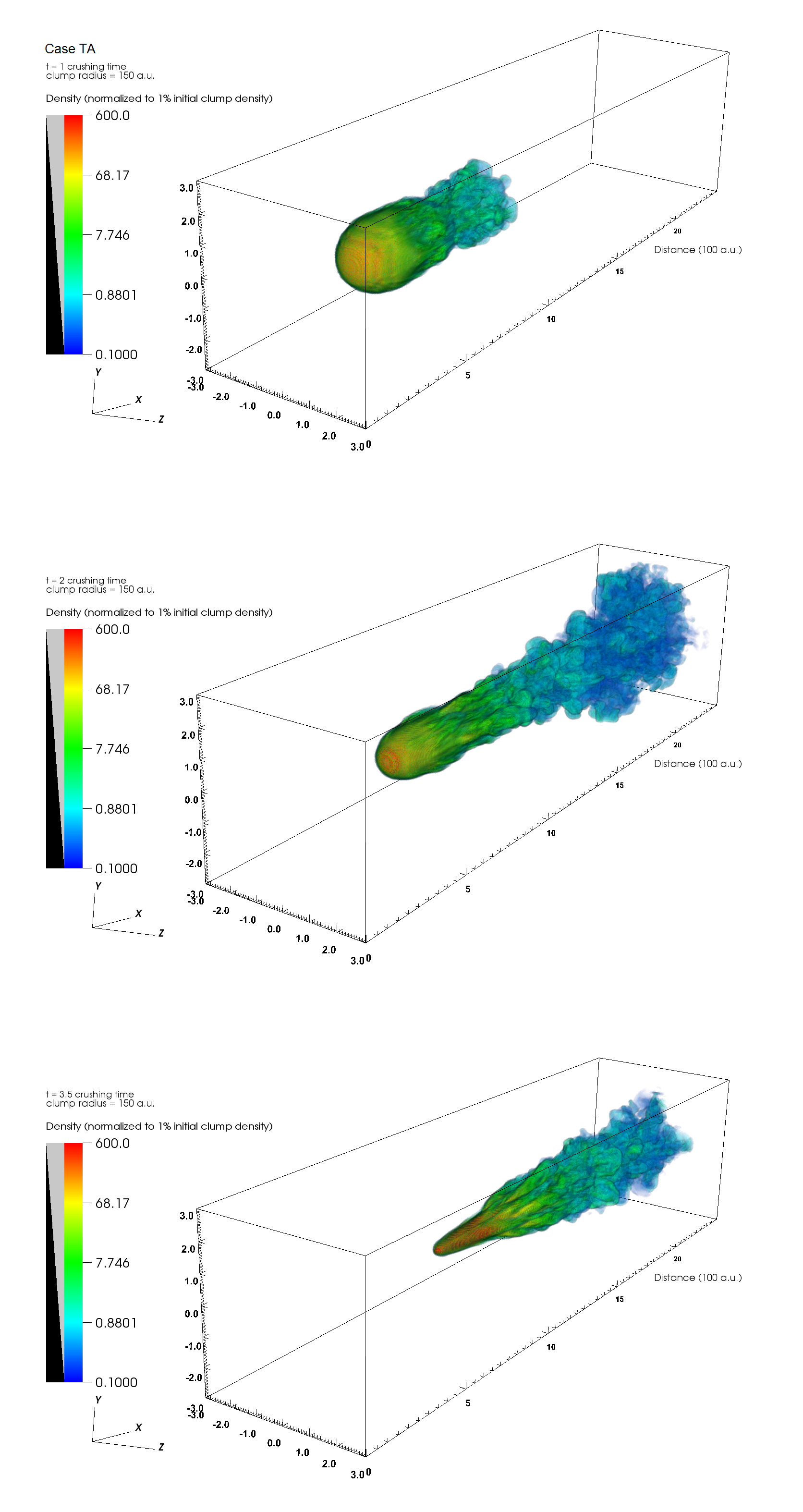}
\caption{Case of strong toroidal only, aligned with shock normal. Evolution of clump material at 1, 2 and 3.5 clump crushing time. The color indicates clump material concentration, normalized by initial value.}
\label{fig03}
\end{figure}

\placefigure{fig03}

\begin{figure}
\includegraphics[scale=.25]{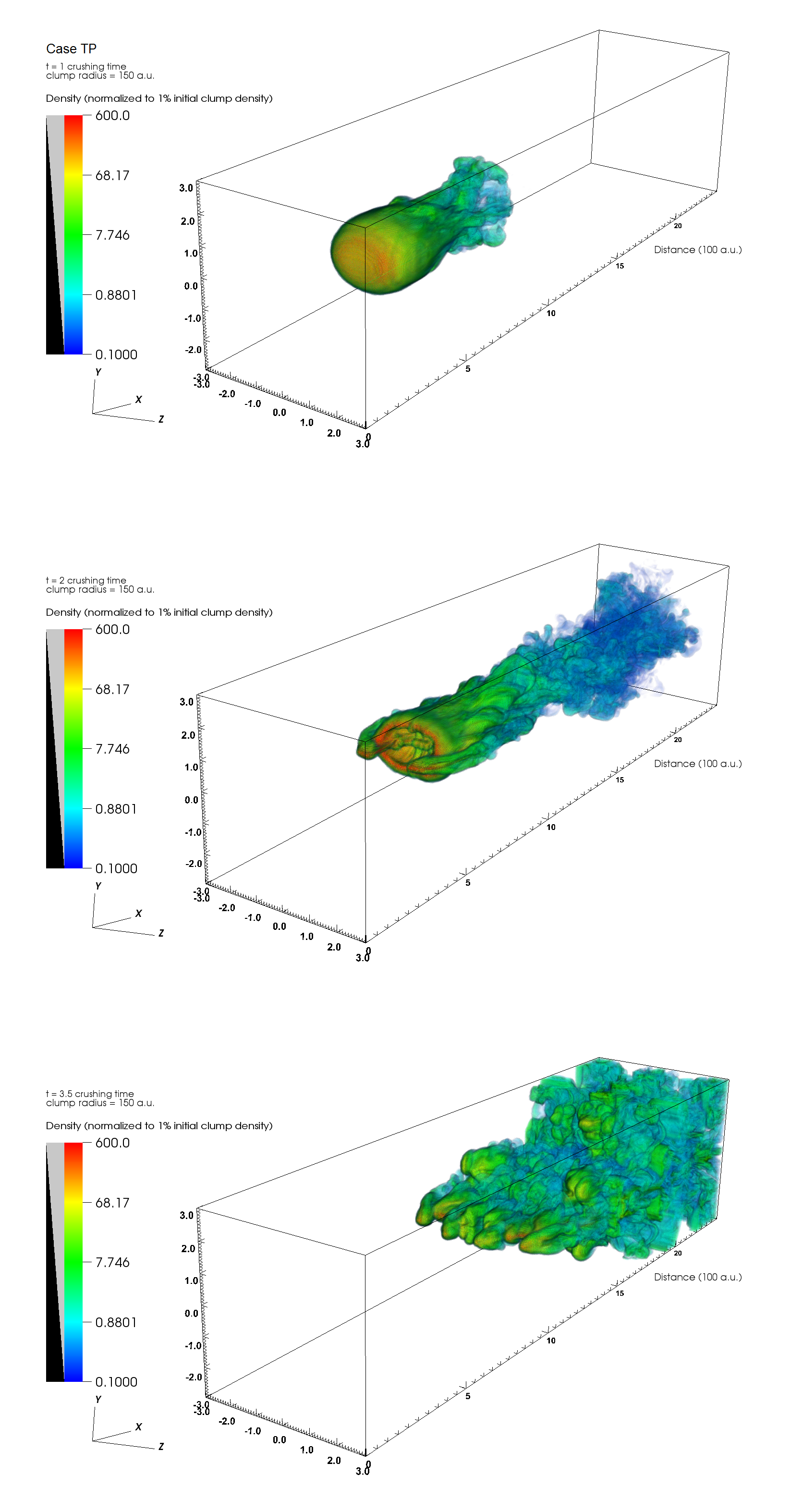}
\caption{Case of strong toroidal only, perpendicular to shock propagation direction. Evolution of clump material at 1, 2 and 3.5 clump crushing time. The color indicates clump material concentration, normalized by initial value.}
\label{fig04}
\end{figure}

\placefigure{fig04}

We now turn to the poloidal strong field cases. Figure 05 shows the simulation of a shocked clump when the internal field is poloidal and aligned with the the shock normal  (case PAS). In this run, there is a strong field concentration of field at the clump axis, as well as a relatively weak field near the clump surface. When the axis is aligned with the shock normal, we can see that during the compression phase $t = \tau_{cc}$, the clump is compressed radially as in the unmagnetized case. Note however that a depression develops along the clump axis as the incident flow's ram pressure is relatively unimpeded there by the magnetic field.  Because the field along the axis is aligned with the flow direction, the evolution resembles the global field parallel case [\citet{mac94}].  However, by $t=2\tau_{cc}$ the differential stresses of internal self-contained poloidal field yield a different evolution compared to both our previous toriodal cases and the uniform field cases.  

While the clump expands laterally as in the unmagnetized case, it then develops a hollow core.  The initial phase of the axial core were already apparent at the earlier times however now we see that the outer regions corresponding to the domains closer to the clump surface with relatively strong magnetic field retain (weaker than the field on the axis, but stronger than the region surrounding the $r_c/3$ point. See Appendix A) their coherence while the incident flow has evacuated the area surrounding the axial core. Thus the poloidal field yields a coherence length associated with the curvature (and tension) of the field around its circumference.  Regions closer to the axis with weak initial field get distorted, compressed and driven downstream while the regions with a strong field or fully flow-aligned field better resist the compression. 

The "shaft" shaped feature surrounded by the hollow core has a relatively low $\beta$ compared to the rest of the clump. It gradually deforms as a result of field line tension (squeezing outwards towards the clump periphery away from the axis) on the timescale of $t=\tau_{B}$, which for these runs is $2.8\tau_{cc}$. Consequently we see at the last frame $t=3.5\tau_{cc}$, that the "shaft" disappears and the clump is fragmented into an array of cold, magnetized "clumplets", similar to the TP case.  
 
Figure 06 shows the simulation with a strong internal poloidal field oriented perpendicular to the shock normal (coded PPS). The influence of the different field orientation is already evident at the first frame $t= \tau_{cc}$.  The initial compression phase has produced an ellipsoidal clump distribution in a similar manner as the toroidal perpendicular simulation (Figure 04).   In this case the internal stresses of the poloidal field change the oriental of the ellipse while also producing substructure due to the smaller scale of field loops ($R\sim 0.5 r_c$ for the poloidal field rather than $R\sim r_c$ for the toroidal case).  By  $t=2\tau_{cc}$ we see a "shaft" and a "ring"  structure develop as in the PAS case, but  now the smaller scale of the loops (radius of curvature) allow these structures to be partially eroded by the incoming shock. The "shaft" is then fragmented by the shock rather than the field pinch, and the "ring" leaves an extended $\bf{U}$-shape structure. As a result, two large clumplets located on the $y-z$ plane form at $3.5\tau_{cc}$. For configurations TA and PP, the initial setup is entirely axisymmetric. 

\begin{figure}
\includegraphics[scale=.25]{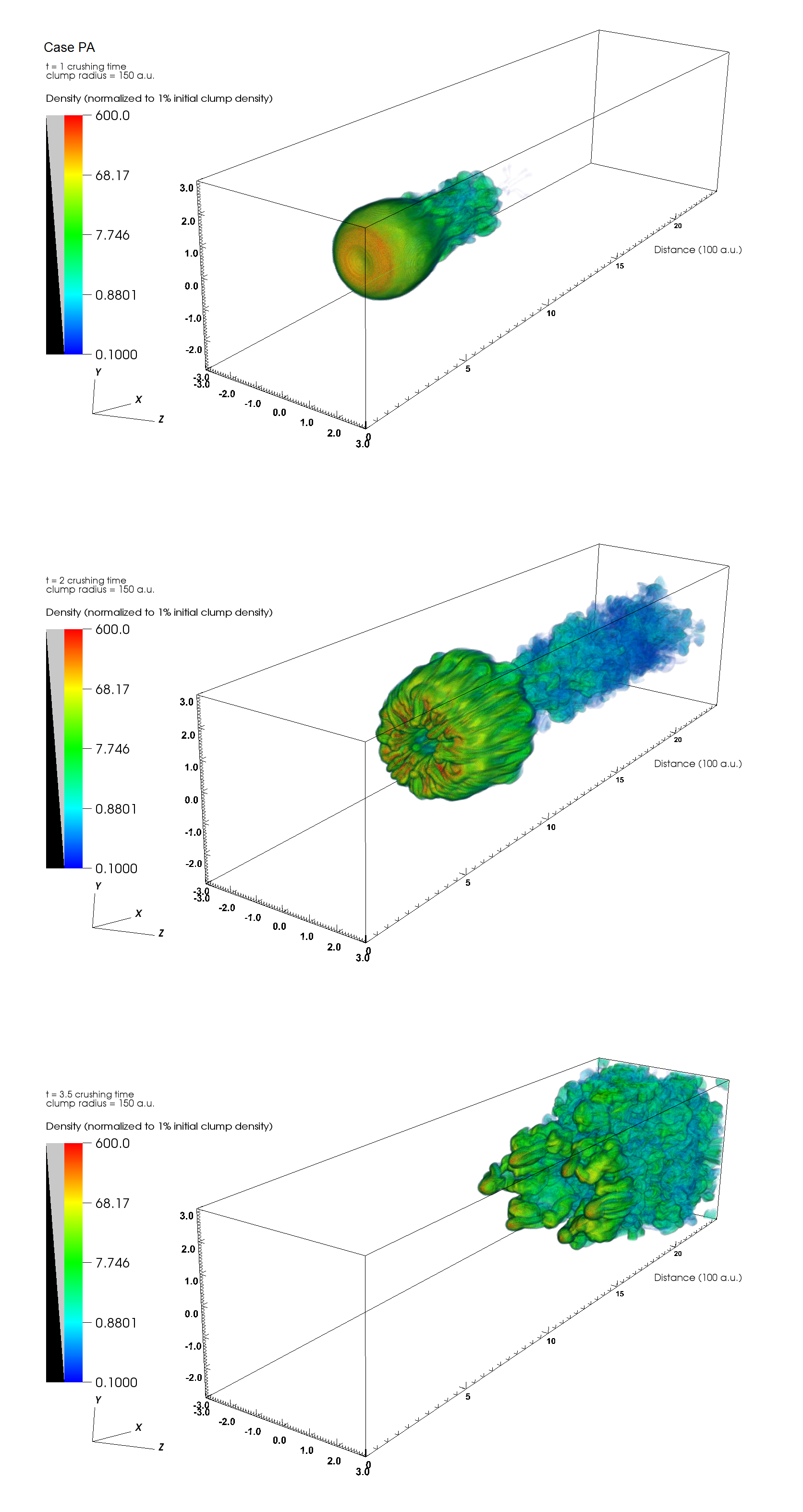}
\caption{Case of strong poloidal only, aligned with shock normal. Evolution of clump material at 1, 2 and 3.5 clump crushing time. The color indicates clump material concentration, normalized by initial value.}
\label{fig05}
\end{figure}

\placefigure{fig05}

\begin{figure}
\includegraphics[scale=.25]{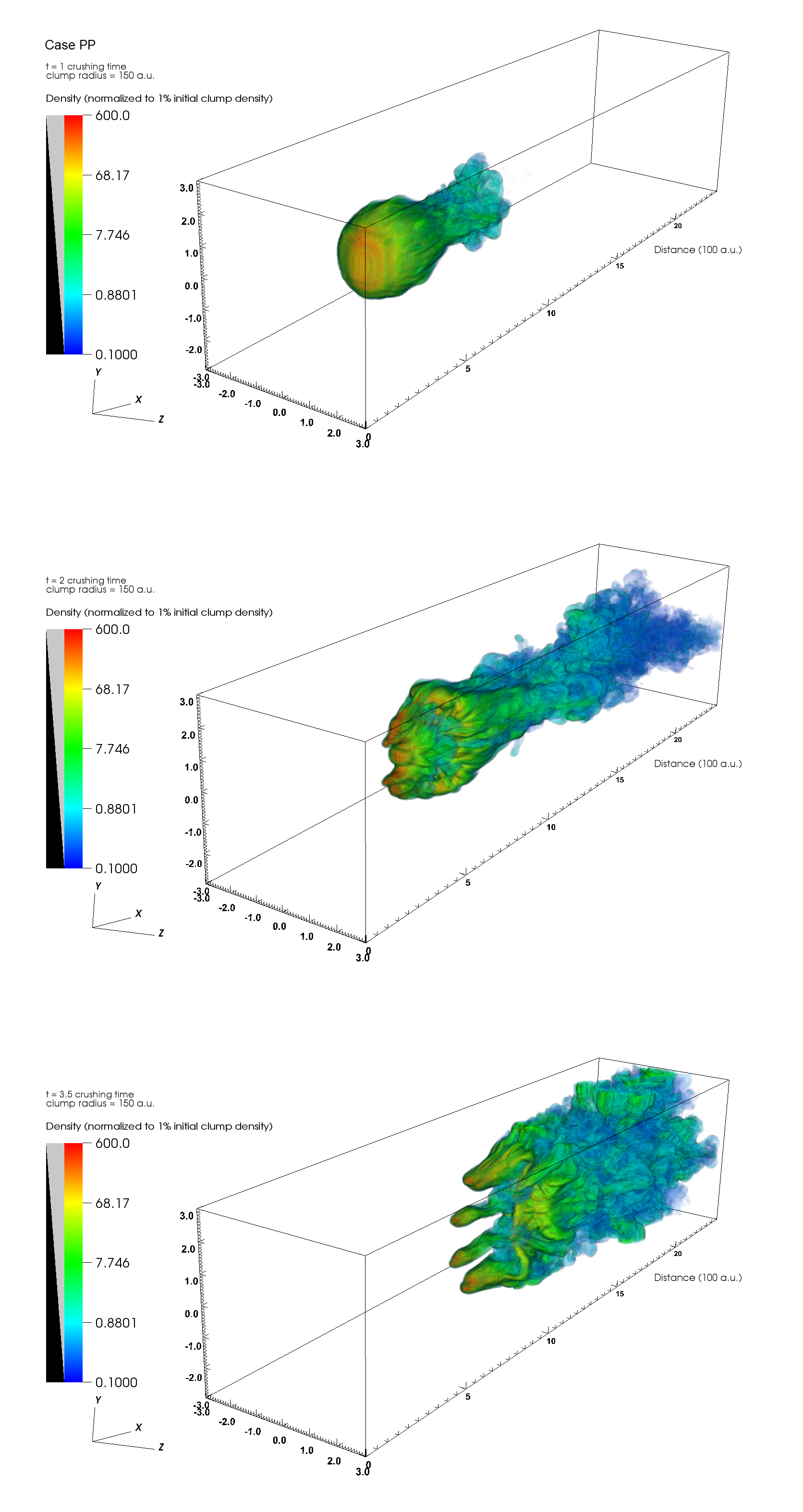}
\caption{Case of strong poloidal only, perpendicular to shock propagation direction. Evolution of clump material at 1, 2 and 3.5 clump crushing time. The color indicates clump material concentration, normalized by initial value.}
\label{fig06}
\end{figure}

\placefigure{fig06}

\subsection{Shocked clumps with a weak self-contained  ordered field}
We now look at the results where the contained magnetic field is relatively weak compared with the previous cases ($\beta_{avg}=1$).  In this regime we still expect to see the field exterting influence over the shock clump evolution but the final outcome on the flow, in terms of global properties, may not sort cleanly between different initial field configurations. 

Figure 07 shows the  simulation of a shocked clump when the internal field is torodial and aligned with the the shock normal  (coded TAW). Here, the most significance difference comparing to the TAS case is that the post-shock clump material does not collapse into a core, instead the ram pressure of the incident flow pushed through the clump axis after the initial compression phase  $\tau_{cc}< t<2\tau_{cc}$. This indicates that the pinch force provided by the toroidal field no longer overwhelms the stresses produced by the flow as it does in the case with stronger initial field and lower initial $\sigma_{B}$. By $3.5\tau_{cc}$, the clump evolves into a series of cold dense clumps as in the hydrodynamic case although the position of the clumps appears to reflect the original toroidal orientation of the field.

Figure 08 shows the case of weak internal toroidal field with its axis perpendicular to the shock normal (coded TPW). Compared to the TPS case in the previous subsection, we can see that the clump opens up at $t=2\tau_{cc}$ similar to the TAW case because of the lack of strong pinch forces. One can still see the the effect of the field in the orientation of the two nascent clumps forming aligned with the z-axis.  Indeed by $3.5\tau_{cc}$, the clump material forms an array of "clumplets" with a stronger distribution along z axis than in $x$ or $y$ which is similar to TPS case. Thus like the TAW case even a weaker self-contained magnetic field still yields an influence over the global flow evolution.

\begin{figure}
\includegraphics[scale=.25]{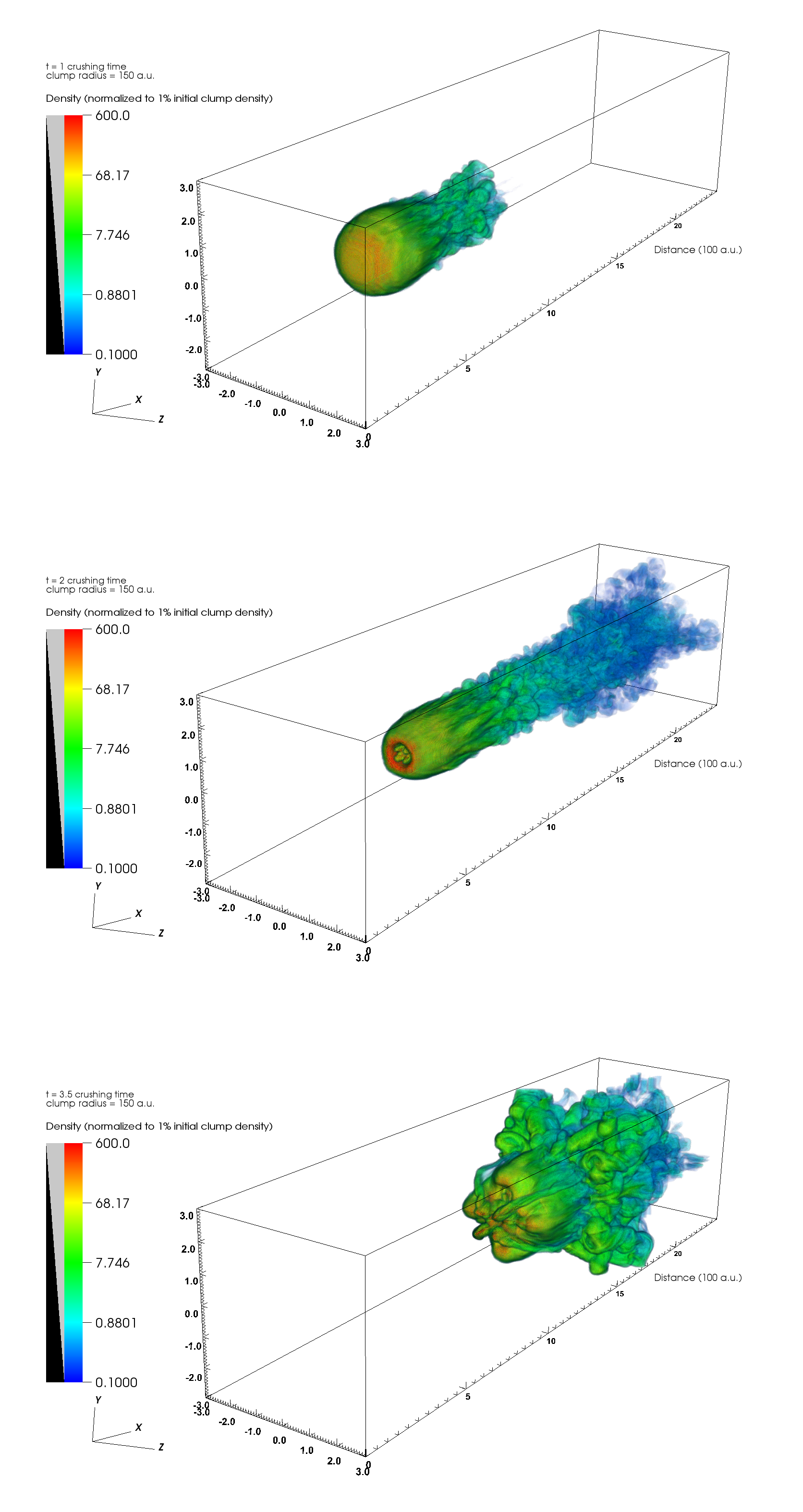}
\caption{Case of weak toroidal only, perpendicular to shock propagation direction. Evolution of clump material at 1, 2 and 3.5 clump crushing time. The color indicates clump material concentration, normalized by initial value.}
\label{fig07}
\end{figure}

\placefigure{fig07}

\begin{figure}
\includegraphics[scale=.25]{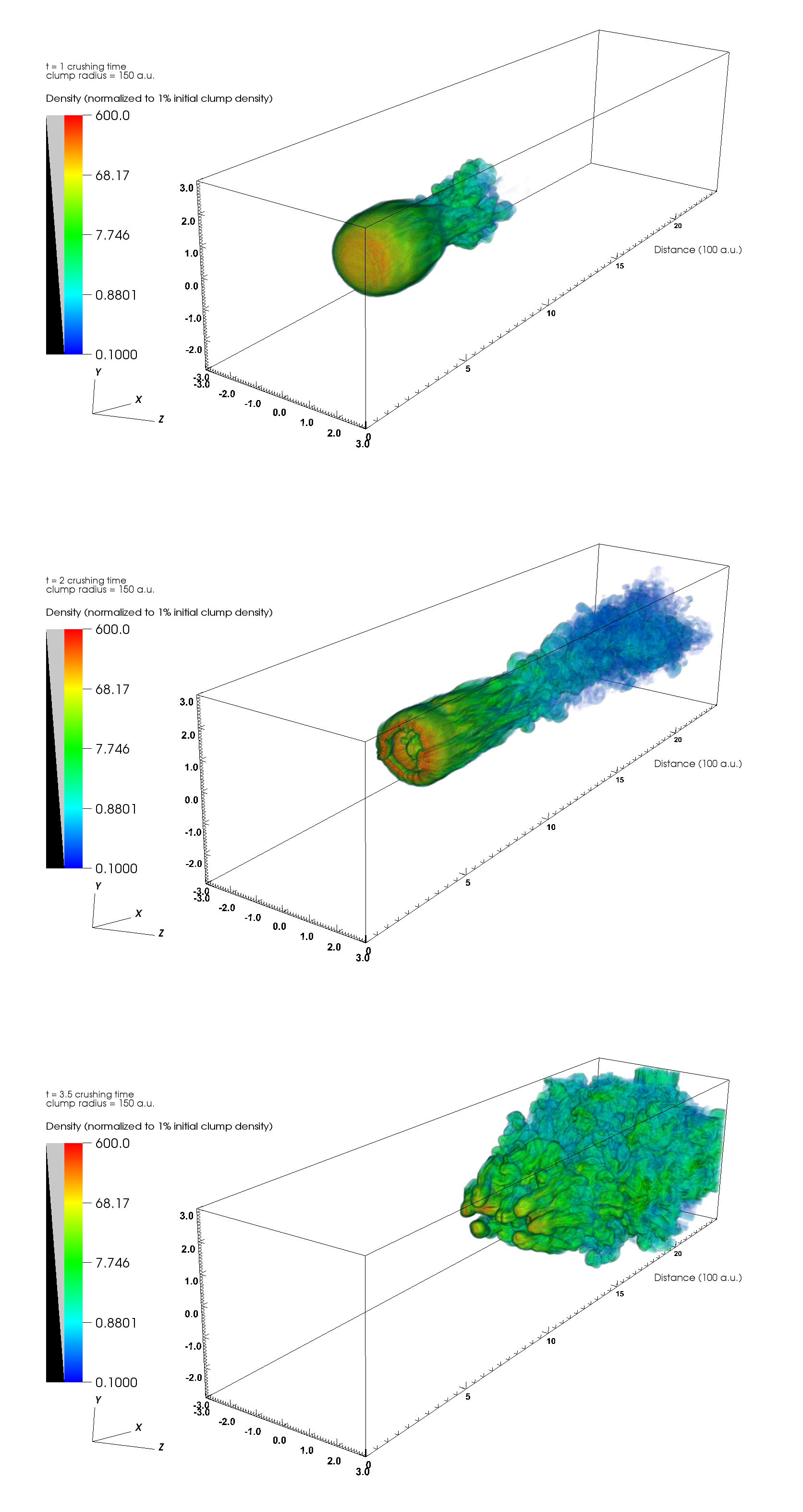}
\caption{Case of weak toroidal only, perpendicular to shock propagation direction. Evolution of clump material at 1, 2 and 3.5 clump crushing time. The color indicates clump material concentration, normalized by initial value.}
\label{fig08}
\end{figure}

\placefigure{fig08}

Figure 09 shows the simulation of a shocked clump when the internal field is poloidal and aligned with the the shock normal  (case PAW). Here the initial morphological evolution is similar to that of the PAS case (Figure 05): at $2\tau_{cc}$, a "shaft" feature is formed, with a "ring" shaped feature surrounding it. By  $3.5\tau_{cc}$, the shaft is destroyed by the internal pinching and the "ring" feature fragments into an array of clumplets due to field pinching and cooling. Notice that the size of the "ring" feature and the spread of the resulting clumplets is smaller compared to the PAS case: an effect that can be attributed to the weaker initial field and its resulting hoop stresses.
 
In Figure 10 we show the simulation with a weak internal poloidal field oriented perpendicular to the shock normal (case PPW). The evolution is comparable with the PPS case. Once again the  $\bf{U}$-shaped feature which forms after the shock has passed through the entire clump is less pronounced due to reduced  pinch forces. Note that we see that the final fragmentation produces two large clumplets at $3.5\tau_{cc}$.

The overall evolution of the weaker field cases shows the effect the field has in terms of the final spectrum of fragments produced by the shock-clump interactions.  Unlike purely hydrodynamic cases the fragmentation of the initial clump into smaller "clumplets" does depend on the the initial field geometry and its orientation relative to the incident shock at least for the evolutionary timescales considered in this study.  Thus even in cases where the field does not dominate the initial energy budget of the clump, the shock dynamics does depend on the details of the initial field.  Note also that in all cases a nearly volume filling turbulent wake develops behind the clump at later evolutionary times. For TA and PP configurations, the initial setup is axisymmetric. But as a result of numerical instabilities and finite domain size, we can observe asymmetry at late frames in Figures 3, 5, 7 and 9.

\begin{figure}
\includegraphics[scale=.25]{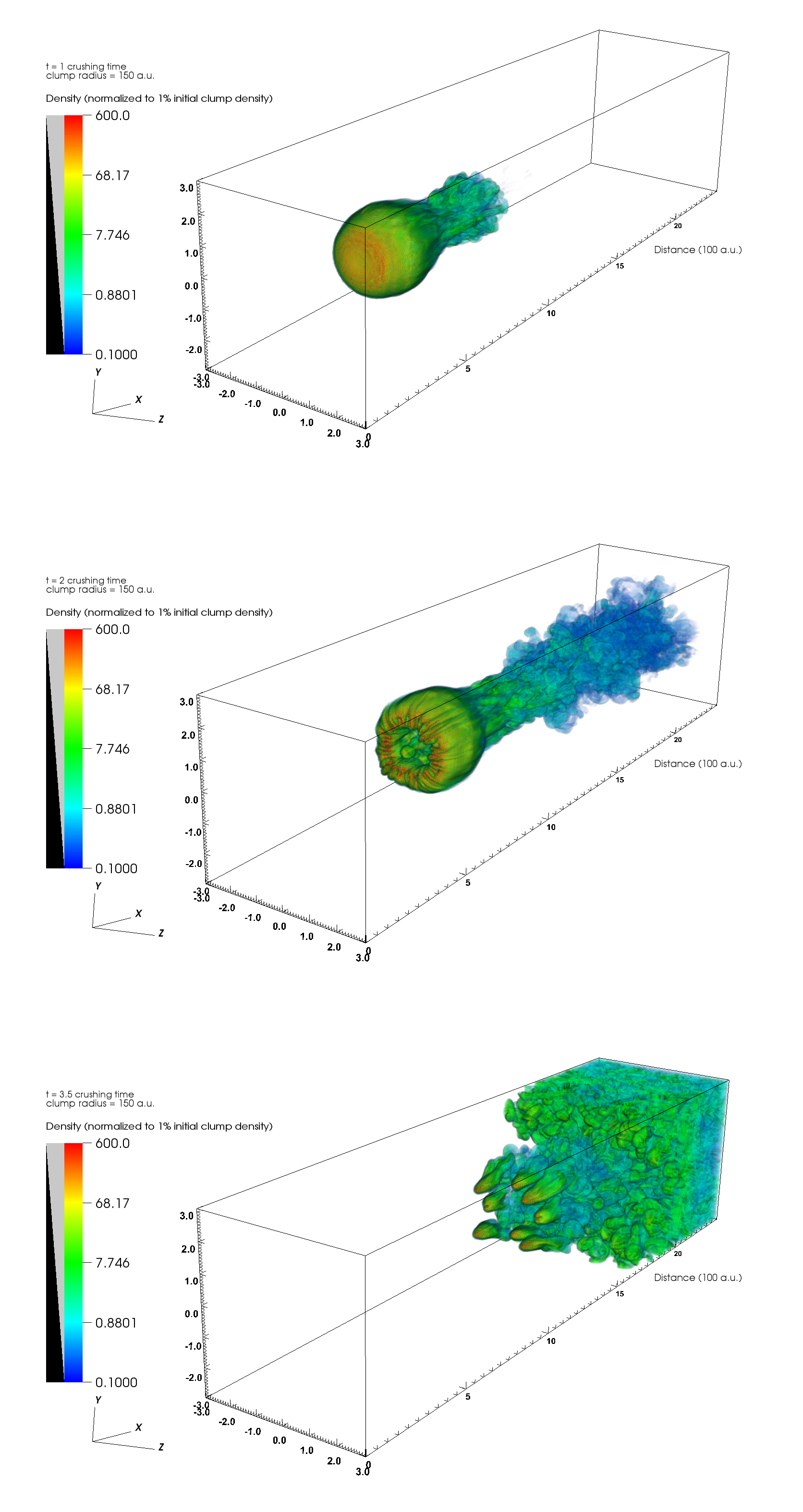}
\caption{Case of weak poloidal only, aligned with shock normal. Evolution of clump material at 1, 2 and 3.5 clump crushing time. The color indicates clump material concentration, normalized by initial value.}
\label{fig09}
\end{figure}

\placefigure{fig09}

\begin{figure}
\includegraphics[scale=.25]{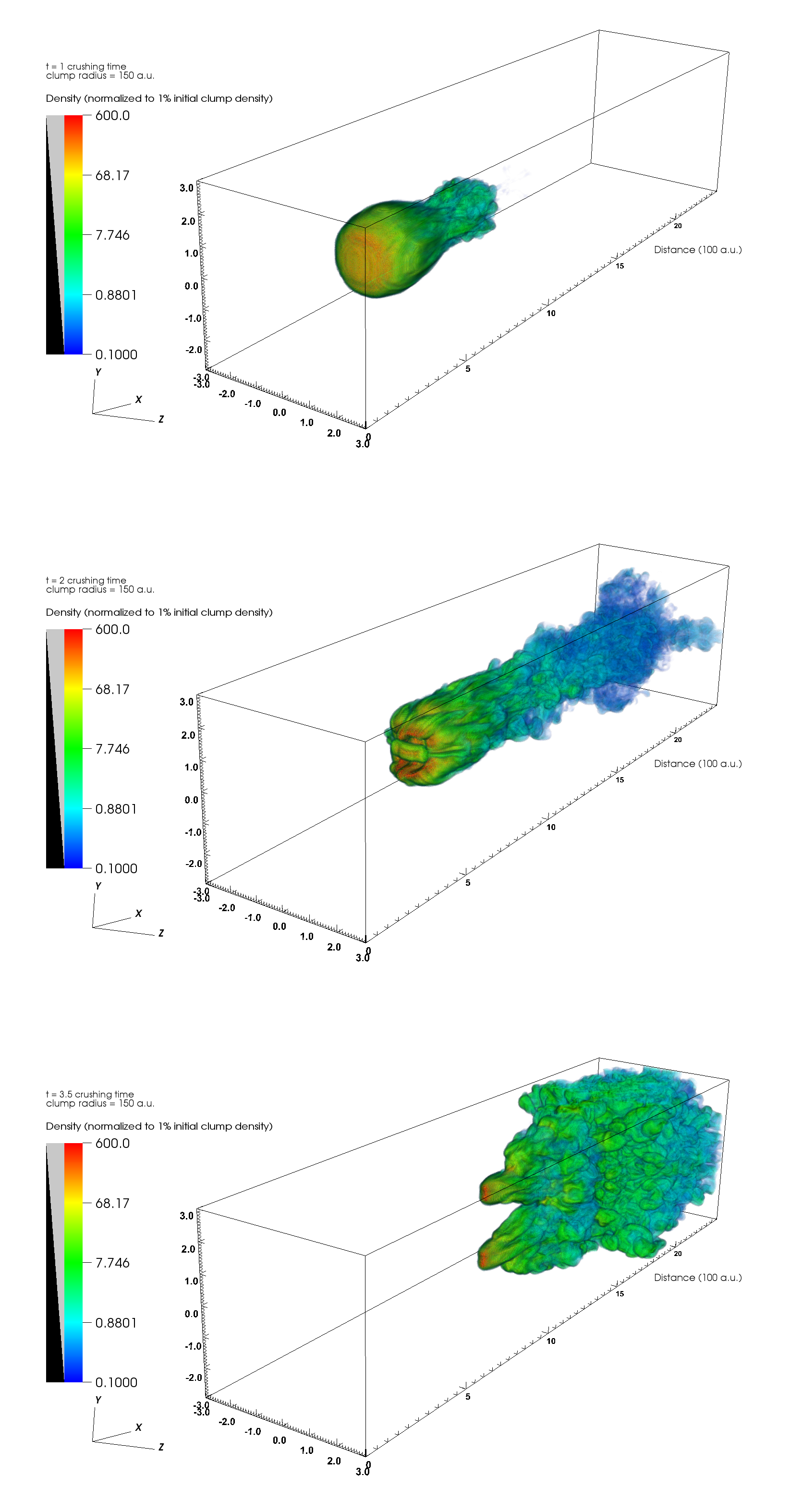}
\caption{Case of weak poloidal only, perpendicular to shock propagation direction. Evolution of clump material at 1, 2 and 3.5 clump crushing time. The color indicates clump material concentration, normalized by initial value.}
\label{fig10}
\end{figure}

\placefigure{fig10}

Magnetic fields can be important in suppressing the instabilities associated with shocked clumps. According to \citet{jon96}, the condition for the magnetic field to suppress the KH instability is that $\beta<1$ for the boundary flows. The condition for the magnetic field to suppress the RT instability is that $\beta<\xi/M=10$. For both strong and weak field cases presented in our paper, the $\beta$ at the boundary flows has a value between $1$ and $10$. Therefore the KH instability is present in all of our cases, shredding the clump boundary flows  and converting them into downstream turbulence. However, even for the weak self-contained field cases, the RT instability is suppressed. To demonstrate, we map the density and $\beta$ (presented by $1/\beta$ in Figure 11) for TAW and PAW cases in Figure 11. We observe that the shocked clump material develops a streamlined shape in both cases. The region where density is concentrated has $1/\beta>  0.1$.

\begin{figure}
\includegraphics[scale=.30]{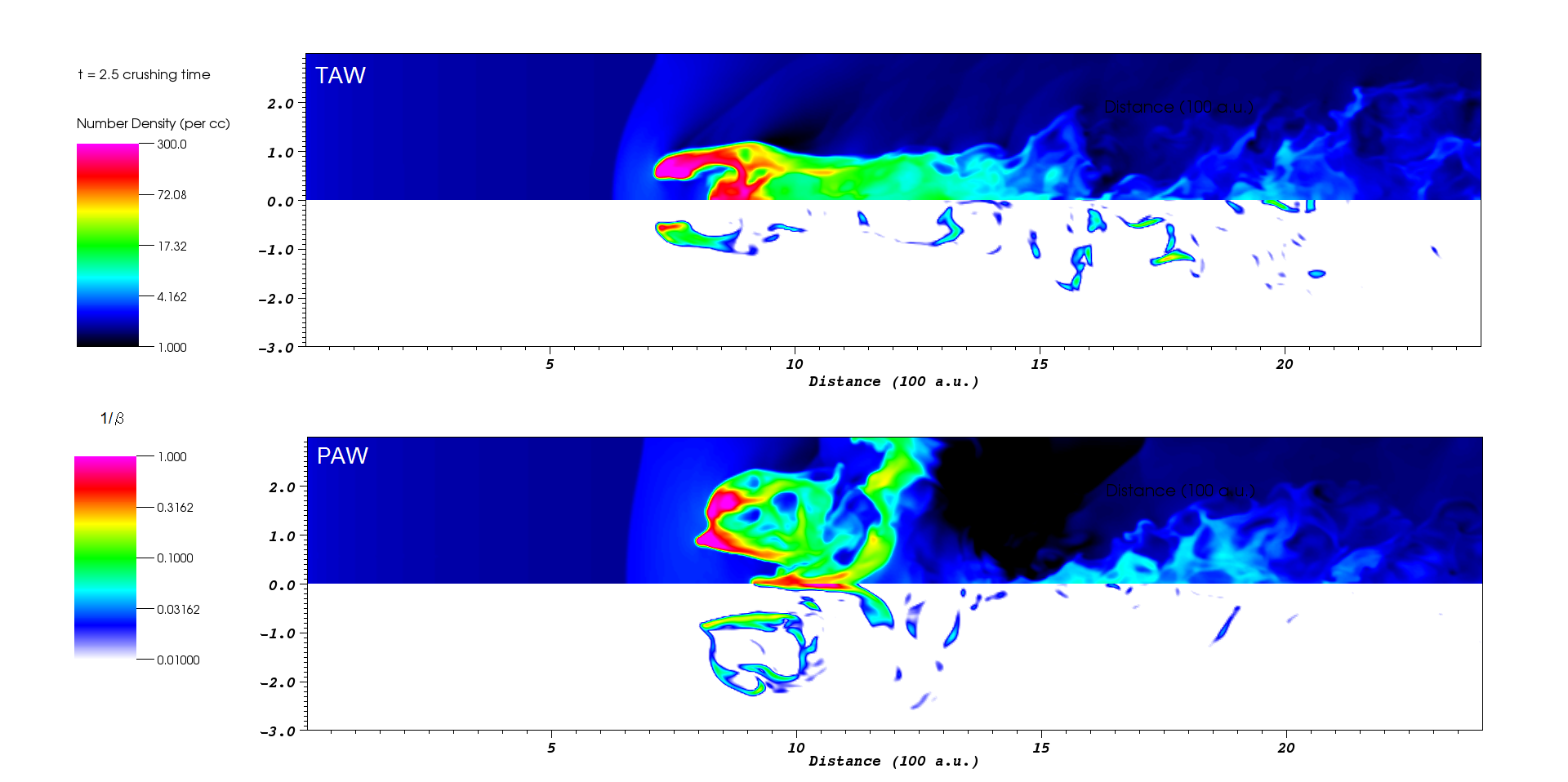}
\caption{Snapshot of shocked clumps cutthrough the center of the domain, at $t=2.5\tau_{cc}$, for the TAW and PAW cases. The upper panel corresponds to the density, the lower panel corresponds to $1/\beta$. }
\label{fig11}
\end{figure}

\placefigure{fig11}

Finally to illustrate the post-shock distribution of magnetic field, we plot the density and field pressure by cutting through the $x-y$ mid plane of the simulation box in Figure 12. It shows that the field follows the clump density distribution, as is expected in our simulations where the diffusion is only numerical and weak.

\begin{figure}
\includegraphics[scale=.30]{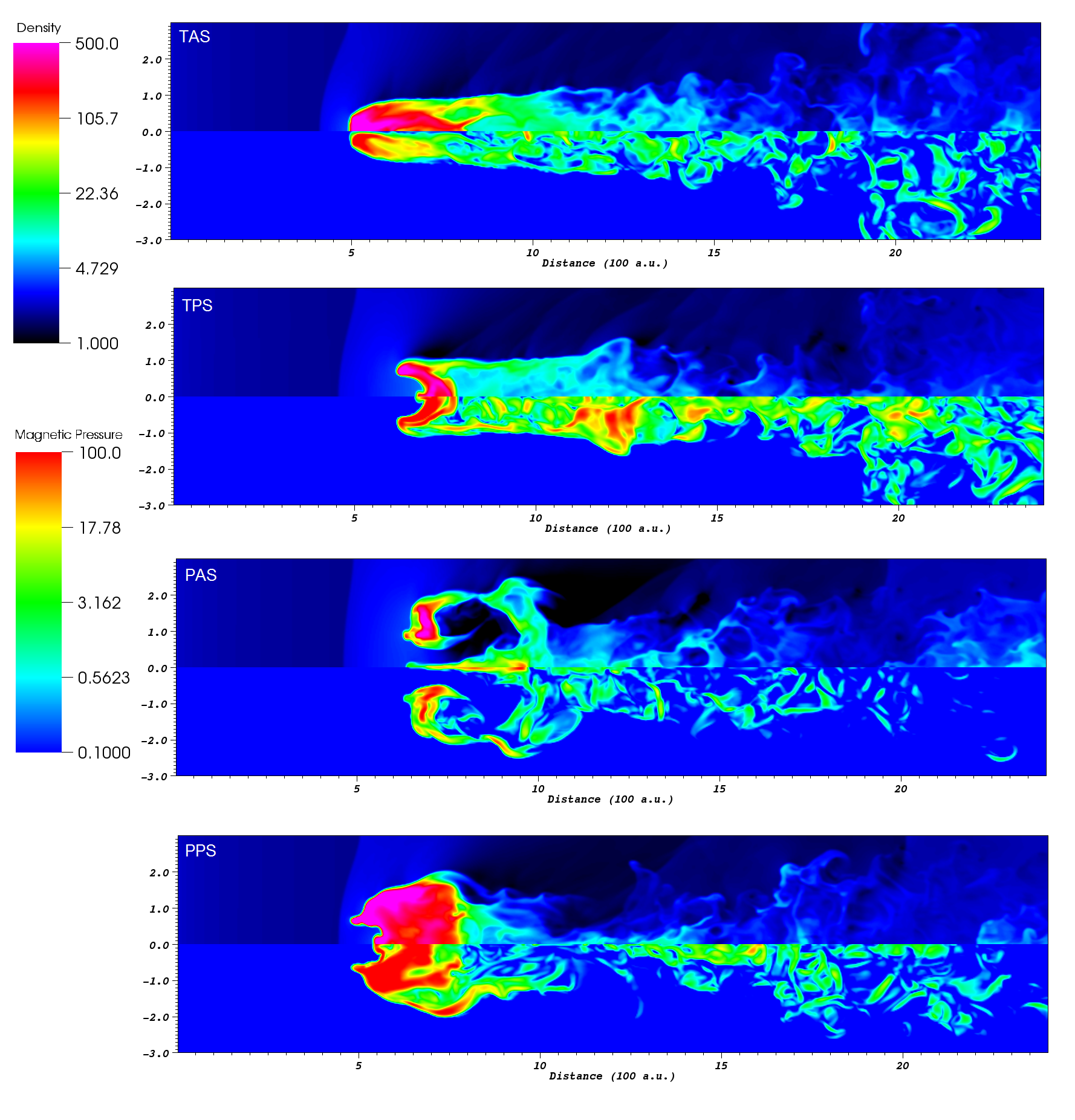}
\caption{Snapshot of shocked clumps cutthrough the center of the domain, at $t=2\tau_{cc}$. The four panels correspond to the TA, TP, PA, PP cases from top to bottom, respectively. The upper half part of each panel shows the clump density, the lower half part shows the magnetic pressure in pseudocolor.}
\label{fig12}
\end{figure}

\placefigure{fig12}

\section{Mathematical Model and Analysis}

\begin{figure}
\includegraphics[scale=.15]{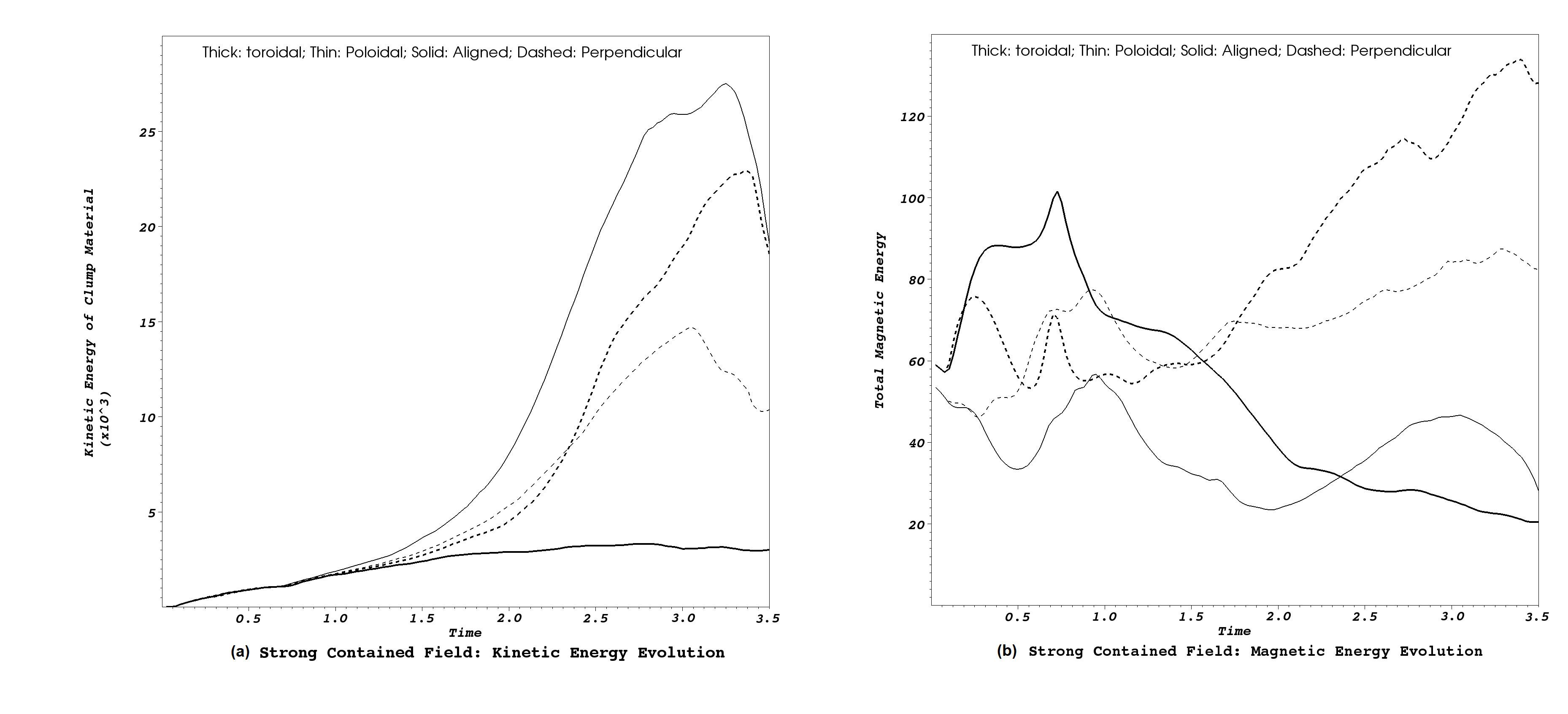}
\caption{Global quantities of the strong self-contained field case: (a) Time evolution of kinetic energy contained in the clump material in computation units, indicating how much energy has transferred from wind into clump. (b) Time evolution of total magnetic energy.}
\label{fig13}
\end{figure}

\placefigure{fig13}

\begin{figure}
\includegraphics[scale=.15]{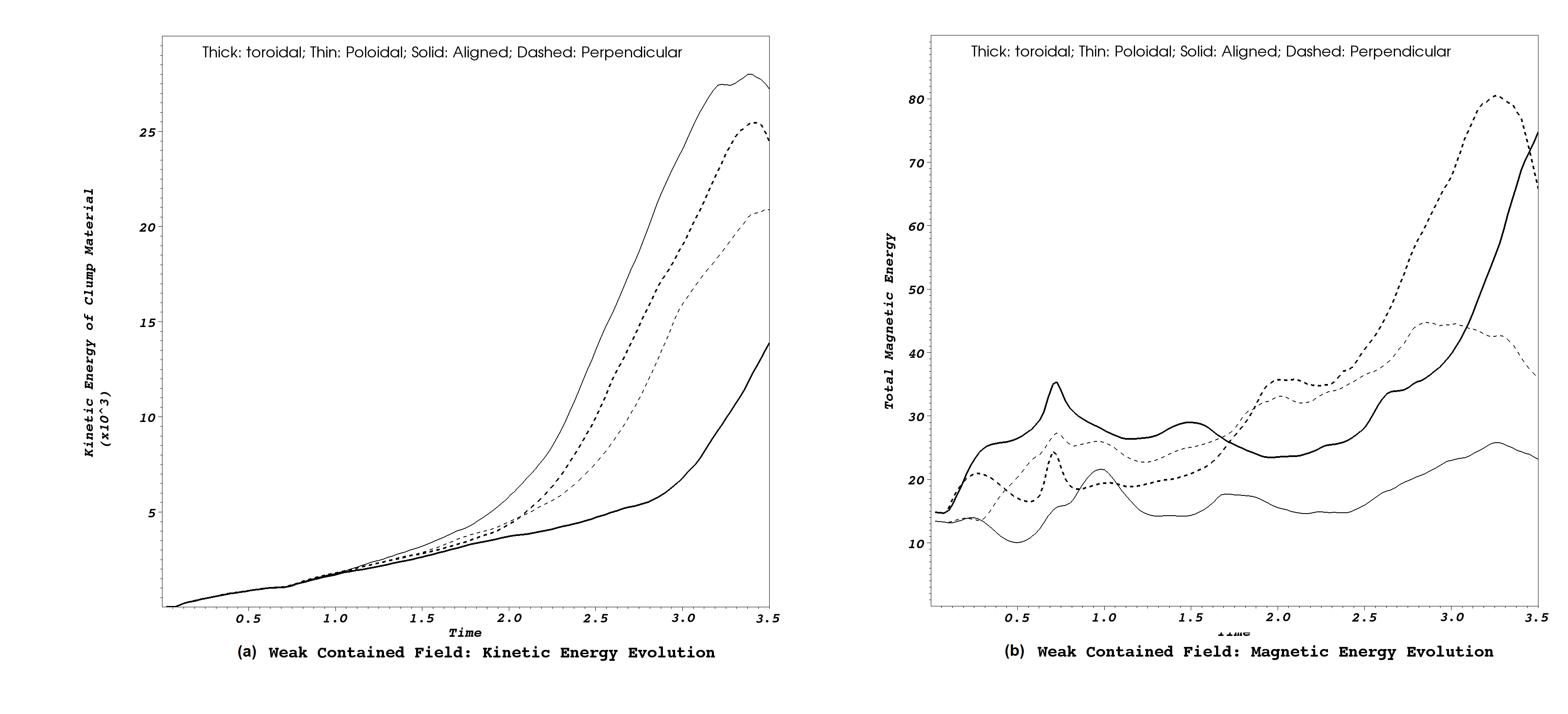}
\caption{Global quantities of the weak self-contained field case: (a) Time evolution of kinetic energy contained in the clump material in computation units, indicating how much energy has transferred from wind into clump. (b) Time evolution of total magnetic energy.}
\label{fig14}
\end{figure}

\placefigure{fig14}

Figure 13(a), (b) show, for the strongly magnetized clump  cases,  the evolution of kinetic energy and total magnetic energy respectively. Figure 14 shows the the analogous  plots  for the weak  field cases. 

In Figure 13(a), we observe that prior to $\tau_{cc}$, the kinetic energy of the clump gained from the incoming shock is similar in all cases. Later, the curves begin to diverge, reach a peak and then descend. The descending feature after $3\tau_{cc}$ is caused by clump material leaving the simulation box. The identical ascending prior to $\tau_{cc}$ and the later diverging behavior for different field configurations will be explained in the subsequent subsection. Similar trend can also be observed for the weak contained field cases of Figure 14(a). 

In Figure 13(b), we observe that the total magnetic energy evolution for the four field configurations are different: TAS case grows and has the highest magnetic energy at $\tau_{cc}$, PAS case fluctuates and has the lowest magnetic energy $\tau_{cc}$.  After $\tau_{cc}$, the TAS curve begins to drop while the other two perpendicular cases continue to rise. At the end of $3\tau_{cc}$, the TPS case has the most magnetic energy, followed by PPS, then PAS. The TAS case dropped to the lowest. In Figure 14(b), the order of contained magnetic energy prior to $\tau_{cc}$ is the same as in Figure 13(a).  However, the TAS curve does not drop afterwards: it continues to rise and at the end of $3\tau_{cc}$, it ranked second in terms of total magnetic energy behind the TPS case. The rest cases have similar feature compared to their strong field counterparts. The magnetic field energy evolution is clearly related to the internal field configuration.

In summary, the kinetic energy transfer and the total  magnetic field variation can be determined by the initial structure of the self-contained magnetic field. To account for the results exemplified in the figures, we propose that the shock-clump interaction incurs two phases, a compression phase and an expansion phase.  

\subsection{Modeling the Compression Phase}

In the evolutionary phase of the shock-clump interaction the transmitted shock passes through the clump and drives it higher densities. After this compression phase energy is then stored in the form of clump thermal pressure and increased magnetic field pressure. During this phase, the kinetic energy of the clump resides  mostly in the form of linear bulk motion and because of the incoming shock, this initial kinetic energy transfer to the clump
is similar  for all of the clump cases we  have considered.  The magnetic energy growth depends on the initial magnetic field  geometry because the shock compression only directly amplifies the field components perpendicular to the shock normal.

We now develop a mathematical model that describes the magnetic field energy for the compression phase. We define $l_{||}$ and $l_{\perp}$ as the thicknesses of the clump along  and perpendicular to the shock normal respectively. We assume that the clumps are initially spherical so initially $l_{\perp,o}=l_x=l_y=l_z=l_{||,o}$ and the shock propagates in the $x$ direction. Subsequently, $l_{||}$ corresponds to the $x$ direction and $l_{\perp}$ refers to the $y$ and $z$ directions,  assuming that the compression is isotropic in the $y-z$ plane.

Assuming that magnetic reconnection is slow on the time scales of the compression phase,  magnetic flux conservation can be used to estimate the magnetic energy increase from compression.  The energy associated with a  uniform field  in the $x-z$ plane  increases $\propto (l_{||}l_\perp)^{-2}$ whereas the energy of a uniform field in the $x$ direction 
will increase $\propto l_\perp^{-4}$.  Then, assuming that the initial field configuration has $\eta B_0^2/8\pi$ stored in the perpendicular component, $(1-\eta) B_0^2/8\pi$ stored in the parallel component, we  obtain the magnetic energy density after compression:
\begin{equation}
\epsilon_B = {B^2\over 8\pi} = {1\over 8\pi}[\eta B_0^2 (2r_c/l_{\perp})^2 (2r_c/l_{||})^2 + (1-\eta) B_0^2 (2r_c/l_{\perp})^4],
\end{equation}
where $r_c$ is the initial clump radius. We use $l_{||,h}$ and $l_{\perp,h}$ to denote the length on the two directions for the case where the clump does not contain any magnetic field, i.e. hydrodynamic case. The magnetic energy density can then be rewritten as:
\begin{equation}
\epsilon_B = (1/8\pi)(\eta B_0^2 (2r_c/l_{||,h})^4  (l_{||,h}/l_{||})^4 (l_{||}/l_{\perp})^2 + (1-\eta) B_0^2 (2r_c/l_{||,h})^4 (l_{||,h}/l_{||})^4  (l_{||}/l_{\perp})^4).
\end{equation}

Assuming that the post compression clump are self-similar (i.e., different in size, but with the same shape) then the ratio of perpendicular and parallel scale lengths is a constant during compression. This allows us to define a constant shape factor $e$, given by
\begin{equation}
e =(l_{||}/l_{\perp})^2 = (l_{||,h}/l_{\perp,h})^2.
\end{equation}
 
  To articulate the influence of the magnetic field compared to a purely hydrodynamic clump we assume that the ratio of the magnetized to unmagnetized clump dimensions in a given direction after compression is inversely  proportional to the ratio of  forces incurred by hydro and magnetized clumps respectively. That is:
\begin{equation}
l_{||,h}/l_{||}=\frac{F-f_B}{F}=1-f_B/F,
\end{equation}
where $F$ is the force exerted by the transmitted shock, and $f_B$ is the "repelling" force exerted by the self-contained magnetic field (see appendix B). The ratio of these two forces is proportional to the magnetic and kinetic  energy densities, that is
\begin{equation}
f_B/F = \frac{\alpha B_0^2}{6\pi \rho_s v_s^2},
\end{equation}
where $\alpha$ is a dimensionless number that depends on the magnetic field configuration, and $\rho_s$ and $v_s$ are the density and velocity behind the transmitted shock. For example, if the  repelling force is from the magnetic pressure gradient  $\nabla P_{B}$ only, and the magnetic field is distributed in a thin shell of radius $r_c/3$, then 
\begin{equation}
f_B = \frac{3}{r_c} \frac{B_0^2}{8 \pi}
\end{equation}
per unit volume. On the other hand, the ram pressure acting on the clump has:
\begin{equation}
F = \frac{\rho_s v_s^2 \pi r_c^2 }{4\pi r_c^3/3}
\end{equation}
per unit volume. Therefore from the above two expressions we obtain that in the case considered, $\alpha = 3$.

Because the self-contained magnetic field is curved with a positive radius of curvature, a magnetic tension force in $\textbf{J} \times \textbf{B}$ is present and can cancel some of the repelling force from the field pressure gradient. For instance, in the toroidal perpendicular case, the  tension force along the $x$ direction is $\partial_x B^2/4\pi$. The tension force therefore reduces $\alpha$ to $\alpha = 1$. We define $\mu$  as the ratio of the initial averaged clump magnetic energy density and the external energy density driving the shock. We also assume $\mu<< 1$ during the compression phase. Specifically,
\begin{equation}
\mu \equiv \frac{B_0^2}{6\pi \rho_s v_s^2} = \frac{2}{3\sigma_{B}} \ll 1.
\end{equation}

We also define the hydrodynamic compression ratio
\begin{equation}
C_h = (2r_c/l_{||,h})^4.
\end{equation}

Combining Eqs (15), (17) and (22), the magnetic energy density after compression can then be written as
\begin{equation}
\epsilon_B = (B_0^2 C_h e/8\pi)(\eta + (1-\eta) e)(1-\alpha \mu)^4.
\end{equation}
Multiplying this total magnetic energy by the volume of the compressed clump gives the total magnetic energy,
\begin{equation}
E_B = (B_0^2 C_h/8\pi)(\eta + (1-\eta) e^2)(1-\alpha \mu)^4 \pi l_{||}^2 l_{\perp} = (B_0^2 C_h l_{||,h}^3/8)(\eta + (1-\eta) e^2)(1-\alpha \mu)^4 (l_{||}/l_{||,h})^3.
\end{equation}
Assuming that all of the different clump field  configuration  cases evolve to  similar shapes after compression (i.e. that $e$ is constant) we then have
\begin{equation}
E_B = eE_{h0} (\eta + (1-\eta) e)(1-\alpha \mu) = E_h (\eta + (1-\eta) e)(1-\alpha \mu),
\end{equation}
where $E_{h0} = B_0^2 C_h l_{||,h}^3/8e$ is the total magnetic field energy in the absence of any  repelling tension force from the self-contained field, and $E_h = eE_{h0}$. Different initial field configurations lead to different strengths of the repelling force and field amplification during the compression and therefore modifying  both $\alpha$ and $\eta$.  Using the strong field case as example, the $\eta$ parameter for the TA, TP, PA, PP cases are $1$, $0.5$, $0.25$ and $0.75$ respectively. From the field gradient and the magnetic tension, we can use $\alpha$  for these four cases: $3$, $1$, $1$ and $3$ (See Appendix B). Using $\mu \approx 0.013$ (from Section 2, $\sigma_{B} \approx 50$) and $e \approx 0.25$ (from the approximated ratio $l_{||}/l_{\perp} \approx 0.5$), we find the total magnetic energy for the TA, TP, PA, PP to be: $0.96E_h$, $0.61E_h$, $0.43E_h$ and $0.78E_h$, respectively. Therefore at the end of the compression phase, the total magnetic energy from high to low is: TA, PP, TP, PA. These theoretically predicted ordering exactly agrees with the line plots  of Figure 13(b) from the simulations.  
 
 The simulations also justify the underlying assumption of Eq.(17), namely that the energy transferred from the shock to the clump material is initially similar in all cases regardless of the  initial field configurations because the field is weak with respect to the impinging flow. This is expressed as
\begin{equation}
(F - f_B) l_{||} \simeq F l_{||,h} 
\end{equation}
and  is evidenced by the  kinetic energy transfer plots Figure 13(a) and Figure 14(a): During the compression phase, all clumps receive identical kinetic energy flux. Note  that our model in the main text ignores  differences in $e$. In the Appendix C we derived the corrections to Eq.(25) when differences in $e$  are allowed.

\subsection{ Expansion Phase}

Unlike the compression phase, in the expansion phase a large fraction of the  kinetic energy of the clump comes from expansion motion parallel to the shock plane.  However the specific evolution of this phase depends on which  two distinct circumstances  arise at the end of the compression phase: Either (1) the magnetic pressure gradient and tension force are small compared to the pressure force exerted by the shock or (2) the magnetic pressure gradient and tension force dominate  over the shock. 

If the shock is still dominant at the end of the compression phase (circumstance 1), the clump will expand similarly to the hydrodynamic case. During this phase, the magnetic field inside the clump acts against this expansion: the clump material is doing work to the self-contained magnetic field (mainly via field stretching) in order to expand. Thus, in general, more magnetic energy at the end of the compression phase means a stronger force opposing the expansion. The kinetic energy in the expansion phase shows differences for the different field configurations: the higher the self-contained field energy at the end of the compression phase, the lower the kinetic energy transfer efficiency in the expansion phase. The  ordering of kinetic energy transfer efficiency in the expansion phase from high to low is then PA, TP, PP, TA. This again exactly agrees with our plots Figure 13(a) and Figure 14(a). 
 
In addition for circumstance (1), the expansion phase also sees a switch in the nature of  field amplification: the field is amplified according to how much kinetic energy is transferred into the expansion motion. Thus the ordering of the magnetic field amplification in the expansion phase will be the same as the ordering for the kinetic energy transfer in that phase. In Figure 14 (b), the weak field cases follow this pattern: the TAW, TPW and PPW curves reverse their ordering when entering the expansion phase, giving them the same ordering as the kinetic energy transfer plot Figure 14(a). The PAW case does not conform with the prediction of the model because most of the field lines are parallel to the shock propagation direction so that they do not get amplified by the stretching from the expansion motion on the $y-z$ plane.

If the shock is no longer dominant at the end of the compression phase (circumstance 2 above), then  the clump evolves under the influence of a significant  Lorentz force. The comparison between the TAS (Figure 13(b)) and TAW (Figure 14(b)) cases exhibits the transition and the distinction between circumstance (1) vs. circumstance (2) evolution: at the end of the compression phase, the TAW case expands while the TAS case shrinks. 

The requirement for these distinct evolutions to arise  can be predicted using a dimensionless ratio calculated from the  parameters of the initial field configuration. Assuming that the pressure from the expansion in the direction perpendicular to the toroidal field lines in a TA case is $1/3$ of the total post shock ram pressure,  the ratio between the total magnetic pressure and the pressure of the expansion motion is given by
\begin{equation}
r_e = \frac{(B_0^2 C_h e/8\pi)(\eta + (1-\eta) e)(1-\alpha \mu)^4}{\rho_s v_s^2 /3}.
\end{equation}

Using the parameters $\alpha = 3$, $\mu_{strong} = 0.013$, $\mu_{weak} = 0.005$, and a compression ratio $C_h = (2R/l_{||,h})^4 \approx 3.5^4 \approx 150$ we find that $r_e \approx 1.42$ for the TAS (circumstance 2) case and $r_e \approx 0.44$ for the TAW case (circumstance 1) respectively. Intuitively the threshold for the toroidal configuration to expand would require $r_e \le 1$. Thus in the TAS case, the field pinch is dominant at the end of compression phase and the clump collapses down to the axis; whereas in the TAW case the expansion is dominant and the clump behaves similar to a hydrodynamic case.

\subsection{Mixing of Clump and Ambient Material}

Figure 15(a), (b) show the mixing ratio of wind and clump material at $\tau_{cc}$ and $3\tau_{cc}$ for the strong field cases. Figure 16(a), (b) show the mixing ratio of wind and clump material at $\tau_{cc}$ and $3\tau_{cc}$ for the weak  field cases. We define a wind-clump mixing ratio  in a single computational cell as
\begin{equation}
\nu = \frac{2 min(n_c, n_w)}{n_c+n_w}, 
\end{equation}
where $n_c$ and $n_w$ denote the clump and wind number densities, respectively. This definition shows that $\nu=1$ means perfect mixing: there is equal number of clump and wind particles in the cell, while $\nu=0$ means no mixing at all. In Figure 15, we see that the mixing ratios for the four strong self-contained field cases are almost identical at early times. This is consistent with the fact that at early times the clump as a whole is in the processes of being accelerated as along the shock propagation direction. The only mixing between clump and wind occurs at the edges of the clump from the interaction with the incoming shock. The strong field prevents strong mixing.

In the weak magnetic field cases, the toroidal configurations do not see a significant increase in the early time mixing ratio compared to the strong field case (Figure 16(a)). This is because the toroidal case has most of its magnetic field concentrated at the edges of the clump (See Appendix A). Thus the average plasma $\beta$ on the outer edge is still small enough to contain the clump material. In the weak poloidal configuration cases however, the magnetic field is concentrated at the center of the clump and accordingly  the PAW and PPW cases have the largest magnetic $\beta$ on the outer edge of clump, making them the most susceptible to early shock erosion. This explains the significant increase we see in the initial mixing ratio in the PAW and PPW cases (Figure 16(a)). 

\begin{figure}
\includegraphics[scale=.35]{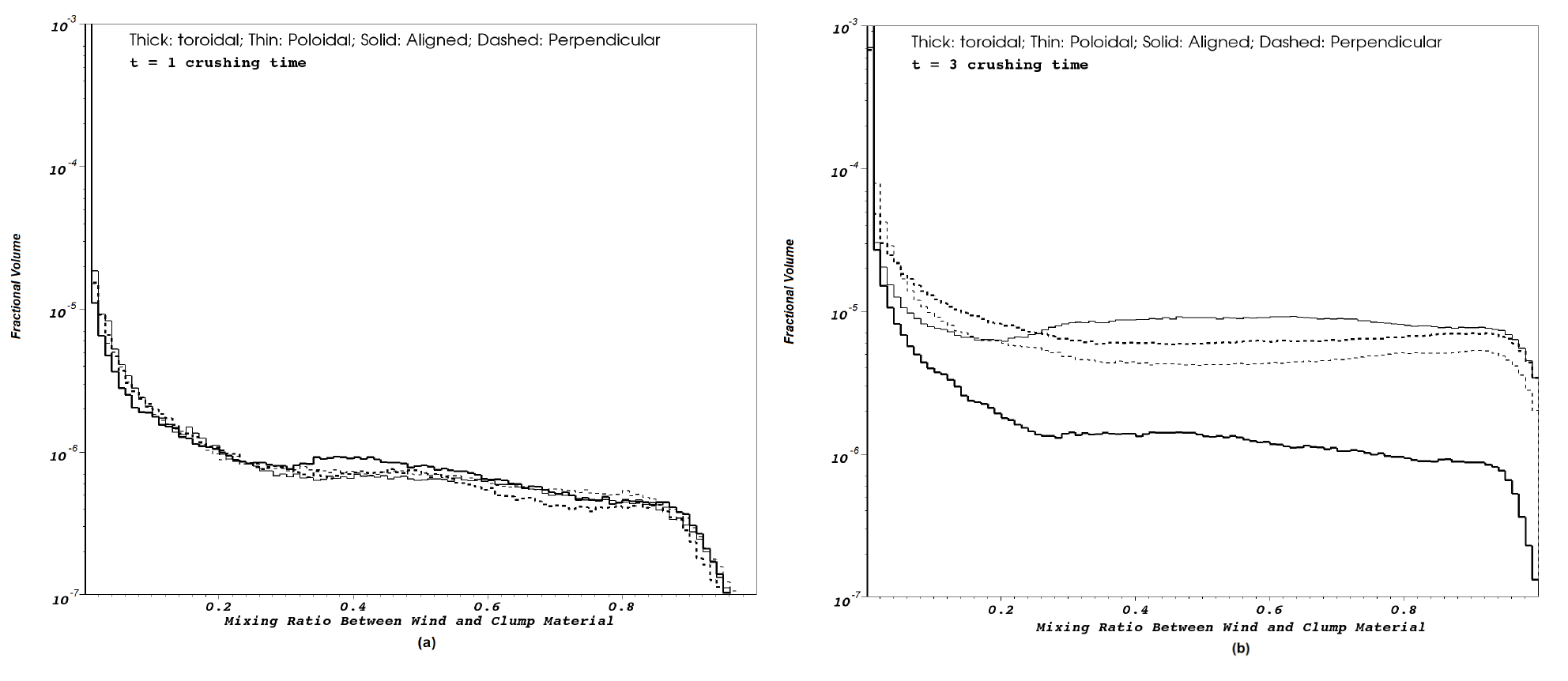}
\caption{Wind-clump mixing ratio for the strong self-contained field case: (a) at $\tau_{cc}$. (b) at $3\tau_{cc}$. The color codings and their corresponding simulations are labeled in the plot}
\label{fig15}
\end{figure}

\placefigure{fig15}

\begin{figure}
\includegraphics[scale=.35]{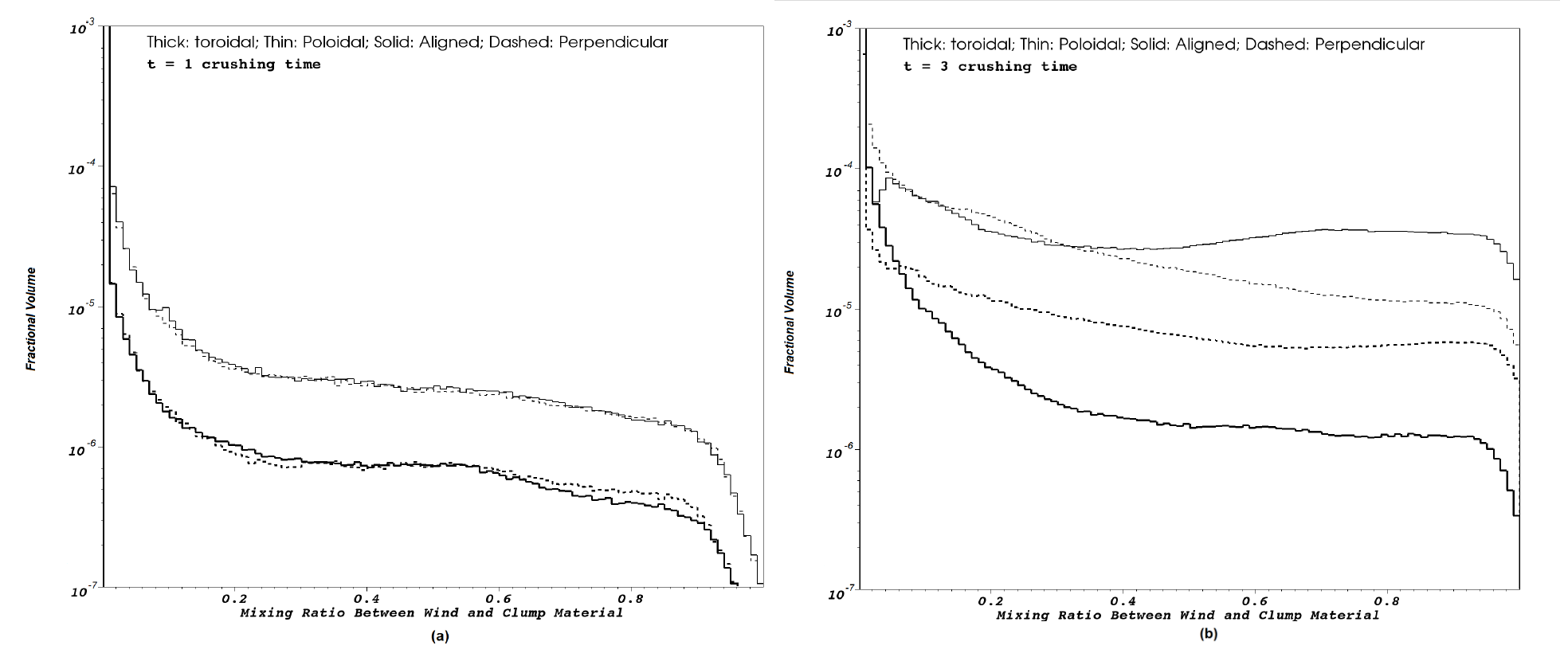}
\caption{Wind-clump mixing ratio for the weak self-contained field case: (a) at $\tau_{cc}$. (b) at $3\tau_{cc}$. The color codings and their corresponding simulations are labeled in the plot}
\label{fig16}
\end{figure}

\placefigure{fig16}

The late mixing ratio depends on how much kinetic energy is transferred from wind to clump. At late times the PA configuration has the highest mixing ratio of the four studied cases. The PP and TP cases  have  intermediate mixing ratios, and the TA has the lowest mixing ratio. This ordering  agrees with the ordering of kinetic energy transfer: the more force resisting compression from the self-contained magnetic field in the early phase the less the kinetic energy transfer occurs in the expansion phase, and the less the mixing. The late mixing ratio also partially depends on the efficacy of enhanced turbulent mixing downstream. The 3-D images in the  previous section Figures 3 to 10, we can identify the downstream turbulence of the TA and PA cases as the least and most volume filling respectively.

\section{Conclusion}

We have studied the evolution of clumps with initially self-contained magnetic fields subject to interaction with a strong shock using both numerical simulations and analytic theory. Our results show a new variety of features  compared to previous work on shock-clump interactions with magnetic fields, which considered only cases in which the field threading the clumps was anchored externally [\citet{jon96}, \citet{gre00}]. 

We find that the evolution of the total magnetic energy and kinetic energy of clumps depends primarily on the relative strength of the self-contained magnetic field, the incoming supersonic bulk kinetic energy (characterized by the $\mu$ parameter) and the geometry of the magnetic field (characterized by the $\eta$ and $\alpha$ parameters). We identified two phases in the clump evolution that we characterized by "compression" and "expansion" phases.

In general, we found strong distinctions in clump evolution  depending on the relative fraction of field in the clump aligned  perpendicular to or parallel to the shock normal. This was demonstrated by considering distinct  field configurations  that we called "toroidal" and "poloidal" and for each case comparing the shock clump interactions when the symmetry axes were aligned with the shock normal and perpendicular to it. The evolution of the clump magnetic fields seen in our simulations can be described by the mathematical model culminating in Eq.(25)  during its compression phase.

The kinetic energy transfer from the supersonic flow to the clumps is similar in the compression phase for all of our cases considered but develop differences in the expansion phase depending on the initial field geometry and orientation, which in turn determines how much field amplification occurs in the compression phase.  The evolution of the clump in the expansion phase depends on whether the shock or the magnetic field is dominant at the end of the compression phase.

The extent to which  clump material mixes with the wind material also depends primarily on the field orientation: in general, the more the initial field is aligned perpendicular to the shock normal, the better the clump can deflect the flow around the clump and the less effective the  mixing. Equivalently, the better aligned the field is with the shock normal, the more effective the clump
material gets penetrated by the incoming supersonic flow, gains kinetic energy in expansion, and enhances mixing.

These simulations may provide morphological links to astrophysical clumpy environments. In our presented study, we use $150 a.u. $ clumps that are typical for young star objects (YSO). However, we also put emphasis on ``weakly cooling" condition that the cooling length as indicated by Eq.(13) is not too small compared to the clump radius. For clumps with much higher clump density, the ratio of clump radius to cooling length $\chi_*$ can be greatly increased. $\chi_*$ can also increase when one tries to scale the simulations to globules that are much larger in size. Therefore in order to gain full understanding of the studied subject, numerical studies that are placed in the parameter regime of ``strongly cooling", where $\chi_*$ is several orders of magnitude greater than its current value, are necessary in the future. Future study may also include more realistic radiative cooling using more recently studied emission lines [\citet{wol94}] and equilibrium heating [\citet{van10}], more realistic internal field geometry, for instance, random field; more realistic multi-physical processes such as thermal conduction, resistivity; and more sophisticated mathematical model.

\acknowledgements{}
Financial support for this project was provided by the Space Telescope Science Institute grants HST-AR-11251.01-A and HST-AR-12128.01-A; by the National Science Foundation under award AST-0807363; by the Department of Energy under award de-sc0001063; by NSF PHY 0903797 and NSF AST 1109285; and by Cornell University grant 41843-7012. The authors would also like to thank Jonathan Carroll and Martin Huarte-Espinosa for useful discussions and suggestions.

\section{Appendix A: Geometrical Setup of Self-contained Magnetic Field}
In order to ensure $\nabla\cdot {\bf B}=0$, the self-contained magnetic field is set up by first choosing  a vector potential distribution, and then taking its curl. The geometry of the toroidal field is best demonstrated using the cylindrical coordinates. The vector potential ${\bf{A}}$ has the following distribution:

\begin{equation}
A_r=0
\end{equation}
\begin{equation}
A_{\theta}=0
\end{equation}
\begin{equation}
A_z=
\begin{cases} B_{0,tor} \frac{f\sqrt{r_c^2-z^2}-r^2}{2fr_c}, &\mbox{ if } r \le f\sqrt{r_c^2-z^2} \\ 
B_{0,tor} \frac{(\sqrt{r_c^2-z^2}-r)^2}{2(1-f)r_c}, &\mbox{ if } r > f\sqrt{r_c^2-z^2}
\end{cases}
\end{equation}
where $B_{0,tor}$ is the desired peak magnetic field intensity, and  $r, \theta, z$ take their usual meanings in a cylindrical coordinate system: $r$ is the distance to the z-axis; $\theta$ is the azimuthal angle; $z$ is the distance to the $x-y$ plane. $f < 1$ is an attenuation factor to cut off the magnetic field when $\sqrt{r^2+z^2}>r_c$, i.e. outside the clump. This vector potential distribution gives the following $\bf{B}$ distribution upon taking the curl:
\begin{equation}
B_r=0
\end{equation}
\begin{equation}
B_{\theta}=
\begin{cases} B_{0,tor} \frac{r}{fr_c},  &\mbox{ if } r \le f\sqrt{r_c^2-z^2} \\ 
B_{0,tor} \frac{\sqrt{r_c^2-z^2}-r}{(1-f)r_c}, &\mbox{ if } r > f\sqrt{r_c^2-z^2}
\end{cases}
\end{equation}
\begin{equation}
B_z=0
\end{equation}
For any given $z$, the magnetic field intensity peaks at $f\sqrt{r_c^2-z^2}$. If $f$ is close to 1, the field will  be concentrated near the outer edge of the clump. In the presented simulations, we take $f=0.9$.

The poloidal field is best demonstrated using the spherical coordinates. It has a vector potential distribution of:
\begin{equation}
A_r = 0
\end{equation}
\begin{equation}
A_{\theta} = -\frac{B_{0,pol}(r_c-r)^2rsin\theta}{2r_c^2}
\end{equation}
\begin{equation}
A_{\phi} = 0
\end{equation}
where $B_{0, pol}$ is the desired peak magnetic field intensity, $r, \theta, \phi$ are the distance to the origin, the polar angle and the azimuthal angle respectively. Notice here $r$ and $\theta$ are defined differently compared to  cylindrical coordinates. The curl of this vector field is:
\begin{equation}
{\bf{B_r}} = 0
\end{equation}
\begin{equation}
{\bf{B_{\theta}}} = 0
\end{equation}
\begin{equation}
{\bf{B_{\phi}}} =  -\frac{B_{0,pol}(r_c-r)(r_c-3r)sin\theta}{r_c}
\end{equation}
We observe  that the magnetic field energy density ${\bf{B}}^2$ peaks at the center $r=0$ and has a weaker secondary maximum at $r=2r_c/3$. The field attenuates to zero at the outer edge of the clump $r=r_c$. There is another zero point in between $r=0$ and $r=r_c$: $r=r_c/3$. The toroidal and poloidal field setup are orthogonal to each other, and can be combined into a more general self-contained magnetic field distribution. The  cases presented in our paper form the basis to understand more complex self-contained magnetic field configurations.

\section{Appendix B: Geometrical Factor of Magnetic Repelling Force $\alpha$}

In the paper, we have worked out the magnetic repelling force for the TA case: 
\begin{equation}
f_B = \frac{3}{r_c} \frac{B_0^2}{8 \pi}
\end{equation}
which gives the parameter $\alpha = 3$. For the TP case, the magnetic tension force is pointing inward with:
\begin{equation}
f_T = \frac{1}{r_c} \frac{B_0^2}{4 \pi}
\end{equation}
assuming the radius of curvature for the magnetic field lines is $R$. This tension force cancels some of the gradient force, which brings $\alpha$ to $1$.

For the PA case, the repelling force from the field gradient remains the same (this is because the average self-contained field pressure is an invariant for the four "strong field" cases). But the curved magnetic field on the outer edge of the clump has an average energy of $B_0^2/2$. The tension force is thus:
\begin{equation}
f_T = \frac{1}{r_c/2} \frac{B_0^2/2}{4 \pi}
\end{equation}
where the field loop's radius of curvature is $r_c/2$. This tension force also brings $\alpha$ down to $\alpha=1$.

For the PPS case, the tension force from the outer edge of the clump can be canceled by the tension force from the center of the clump so that their net contribution to the total repelling force is zero. Therefore we get roughly the same $\alpha$ as in the TA case. 

\section{Appendix C: Correction in the Shape Factor $e$}

In deriving Eq.(25), we used an assumption that no matter what the self-contained field configuration is, the clump is always compressed to a self similar shape if the hydrodynamic setup is unchanged. However, we know that when the self-contained field is ordered, the force it exerts on the clump is inhomogeneous depending on the geometry. The difference in the repelling force therefore results in a difference in the shape factor $e$ introduced in Section 5.1. We now look at how large this correction is for the four studied simulations.

Let us go back to Eq.(15). Assuming the force exerted by the shock on the clump is different on the perpendicular and parallel directions: the force on the perpendicular direction is only a portion of that on the parallel direction, and this portion is fixed for all the cases with the same hydrodynamic setup:
\begin{equation}
F_y = \gamma F_x
\end{equation}
where $\gamma$ is fixed. Then following the same procedure as in Section 5.1, we have:
\begin{equation}
\epsilon_B = E_h(\eta (1-\alpha_x \mu) + (1-\eta) \frac{(1-\alpha_y \mu/\gamma)^2}{1-\alpha_x \mu})
\end{equation}
where $\alpha_x$ and $\alpha_y$ denote different repelling forces from the self-contained field on the x and y direction. 

As in Section 5.1, $\alpha_x$  for the simulated cases TAS, TPS, PAS, PPS are $3$, $1$, $1$ and $3$. Since the perpendicular $\alpha_x$ is just the aligned $\alpha_y$ and vice versa, we know that the $\alpha_y$ for these four cases are $1$, $3$, $3$ and $1$. We use the same parameters as in Section 5.1: $\mu = 0.013$. We assume the incoming shock engulf a spherical sector of the clump with a cone angle $2\theta_e$. Then the compression force applied on the y direction is a fraction of that of the initial incoming shock. This fraction is $\frac{2}{\pi}\int_0^{\theta_e} \frac{1}{2} sin^2 2\theta d\theta$. During the compression process, $\theta_e$  varies from $0$ to $\pi/2$. Therefore we can estimate $\gamma$ as:
\begin{equation}
\gamma = \frac{2}{\pi/2} \int_0^{\pi/2}\frac{\int_0^{\theta_e} \frac{1}{2} sin^2 2\theta d\theta}{\pi/2}d\theta_e = 0.125
\end{equation}
where the inner integration calculates the ratio of average pressure applied on the perpendicular direction when the compressed part of the clump is a spherical cone with cone angle $\theta_e$; the outer integration calculates the average over the compression process where $\theta_e$ varies from $0$ to $\pi/2$. The factor $2$ results from the fact that the perpendicular compression happens on both $+y$ and $-y$ directions.

We can calculate the corrected compressed magnetic field energy for the TAS, TPS, PAS and PPS cases. The results are $0.96E_h$, $0.73E_h$, $0.6E_h$ and $0.93E_h$, for the TAS, TPS, PAS, PPS cases respectively. Comparing to the results presented in Section 5.1:  $0.96E_h$, $0.61E_h$, $0.43E_h$ and $0.78E_h$ for the four cases, we find there is a positive correction to the cases with $\eta < 1$. The ordering of the field amplification factor remains unchanged. Further sophisticated modeling is possible by taking into consideration the dependence of the Lorentz force on the compression ratio: the further the compression, the smaller the magnetic field length scale thus the stronger the repelling force. This results in a model with an integral equation, on which we did not discuss in this paper.

\clearpage

\clearpage

\begin{table}
\caption{Simulation Setups}
\label{tab01}
\begin{tabular}{|c|c|c|c|c|}
\tableline
Code & $\beta_{avg}$ & Field Configuration & Field Orientation (related to shock normal) & $\eta$ \\
TAS & $0.25$ & $ toroidal $ & aligned & 1 \\
TPS & $0.25$ & $ toroidal $ & perpendicular & 0.5 \\
PAS & $0.25$    & $ poloidal $ & aligned & 0.25 \\
PPS  & $0.25$   & $ poloidal $ & perpendicular & 0.75 \\
TAW & $1$ & $ toroidal $ & aligned & 1 \\
TPW & $1$ & $ toroidal $ & perpendicular & 0.5 \\
PAW & $1$    & $ poloidal $ & aligned & 0.25 \\
PPW  & $1$   & $ poloidal $ & perpendicular & 0.75 \\

\tableline
\end{tabular}\\
\end{table}

\end{document}